\documentclass[pra,twocolumn,byrevtex,showpacs,superscriptaddress]{revtex4}

\usepackage{graphicx}
\usepackage{dcolumn}
\usepackage{amsmath}
\usepackage{epsfig}
\usepackage{bm}
\usepackage{amssymb}
\usepackage{hyperref}
\hypersetup{colorlinks=true,linkcolor=blue,citecolor=blue,urlcolor=black}

\begin{document}

\title{Maximum-confidence discrimination among symmetric qudit states}  

\author{O. Jim\'{e}nez}
\affiliation{Departamento de F\'isica, Facultad de Ciencias B\'asicas, Universidad de Antofagasta, Casilla 170, Antofagasta, Chile}
\affiliation{Center for Optics and Photonics, Universidad de Concepci\'on, Casilla 4016, Concepci\'on, Chile}

\author{M. A. Sol\'is-Prosser}

\author{A. Delgado}

\author{L. Neves}
\email{leonardo.neves@cefop.udec.cl} 

\affiliation{Center for Optics and Photonics, Universidad de Concepci\'on, Casilla 4016, Concepci\'on, Chile}
\affiliation{MSI-Nucleus on Advanced Optics, Universidad de Concepci\'on, Casilla 160-C, Concepci\'on, Chile}
\affiliation{Departamento de F\'isica, Universidad de Concepci\'on, Casilla 160-C, Concepci\'on, Chile}

\date{\today}

\begin{abstract}
We study the maximum-confidence (MC) measurement strategy for discriminating among nonorthogonal symmetric qudit states. Restricting to linearly dependent and equally likely pure states, we find the optimal positive operator valued measure (POVM) that maximizes our confidence in identifying each state in the set and minimizes the probability of obtaining inconclusive results. The physical realization of this POVM is completely determined and it is shown that after an inconclusive outcome, the input states may be mapped into a new set of equiprobable symmetric states, restricted, however, to a subspace of the original qudit Hilbert space. By applying the MC measurement again onto this new set, we can still gain some information about the input states, although with less confidence than before. This leads us to introduce the concept of \emph{sequential maximum-confidence} (SMC) measurements, where the optimized MC strategy is iterated in as many stages as allowed by the input set, until no further information can be extracted from an inconclusive result. Within each stage of this measurement our confidence in identifying the input states is the highest possible, although it decreases from one stage to the next. In addition, the more stages we accomplish within the maximum allowed, the higher will be the probability of correct identification. We will discuss an explicit example of the optimal SMC measurement applied in the discrimination among four symmetric qutrit states and propose an optical network to implement it. 
\end{abstract}


\maketitle

\section{Introduction}
The understanding that a quantum system carries information encoded in its quantum state and, thereby, could be used to accomplish information processing tasks gave rise to the fields of quantum information and computation \cite{NielsenBook,BarnettBook}. In order to accomplish such tasks, in the final step one has to read out the previously processed information, which corresponds to determine the final state of the system by measuring it. However, as the quantum state is not itself an observable \cite{PeresBook}, it is not possible to determine it through a single shot measurement, unless it belongs to a \emph{known} set of states which are mutually orthogonal. When this is not the case, i.e., the possible final states are not orthogonal, they cannot be, deterministically, discriminated with certainty and without error even if they belong to a known set. This has led to the development of the area known as quantum-state discrimination (QSD) \cite{Chefles00,Barnett01-2,BergouBook,CheflesBook,Bergou07,Barnett09-2,Bergou10}, where a measurement strategy is devised in order to discriminate \emph{optimally}, according to some figure of merit, among nonorthogonal states.

Despite that, originally, the QSD problem has been introduced in the context of quantum detection (or decision) theory \cite{HelstromBook,Holevo73,Yuen75} long before the birth of quantum information and computation, it quickly became a fundamental tool for these fields. For instance, there is an intimate connection among QSD and probabilistic protocols, like entanglement concentration \cite{Chefles97,Chefles98-1,Croke08,Yang09}, cloning \cite{Duan98,Jimenez10}, and some quantum algorithms  \cite{Bergou03}. Also, there is a connection among QSD and probabilistic realizations of quantum communication protocols like teleportation \cite{Roa03}, entanglement swapping \cite{Delgado05}, and superdense coding \cite{Pati05}. Finally, the use of nonorthogonal states, and, consequently, the impossibility of perfectly discriminating among them, underlies the security in some quantum key distribution protocols \cite{Bennett84}. 

The problem addressed in QSD can be briefly posed as follows. A quantum system is prepared in one of $N$  possible states in the set $\{\hat{\rho}_j\}$, with associated \emph{a priori} probabilities $\{p_j\}$ ($\sum_{j=0}^{N-1}p_j=1$). Both the set of states and the prior probabilities are \emph{known} in advance \cite{CommentUnk}. As the states are nonorthogonal, one has to design a measurement strategy which determine optimally which one was actually prepared.  The optimality criteria are related with some figure of merit, mathematically formulated, and each figure corresponds to a different strategy. The oldest and, perhaps, simplest criterion comprises a measurement that minimizes the probability of making an error in identifying the state \cite{HelstromBook,Holevo73}. For this so-called minimum error strategy (ME), the  necessary and sufficient conditions that must be satisfied by the operators describing the optimized measurement are well known \cite{Holevo73,Yuen75}. Nevertheless, only for a few special cases the explicit form of such measurements have been found \cite{HelstromBook,Ban97,Barnett01-1,Chou03,Andersson02,Yuen75}. A second strategy, first proposed by Ivanovic \cite{Ivanovic87}, allows one to identify each state in the set without error but with the possibility of obtaining an inconclusive result. This strategy, called unambiguous discrimination (UD), is optimized by a measurement that minimizes the probability of inconclusive results. Restricting to pure states, the optimal UD problem was completely solved for the case of two states \cite{Ivanovic87,Dieks88,Jaeger95}, while for more than two states only few analytical solutions have been derived \cite{Peres98,Chefles98-2,Sun01,Pang09}. In this latter case, Chefles \cite{Chefles98-1} showed that UD is applicable only to linearly independent sets.

There exist many other measurement strategies for QSD that optimize differently formulated figures of merit. A discussion about them is beyond the scope of the present work \cite{CommentReview}. Here, we will focus on the recently introduced maximum-confidence (MC) strategy, which is an optimized measurement whose outcome leads us to identify a given state in the set with the maximum possible confidence \cite{Croke06}. The MC measurement can be applied to both linearly independent and linearly dependent states and, unlike the previously discussed ME and UD, it allows a closed form solution for the operators describing the optimized measurement for an arbitrary set of states. In fact, MC encompasses both UD and ME strategies \cite{Bergou10,Barnett09-2}: for linearly independent states it can reduce to UD, where our confidence in identifying the states becomes unity. On the other hand, when the maximum confidence is the same for all states and there is no inconclusive result, MC and ME coincide.

In the original proposal of MC measurement, Croke \emph{et al.} \cite{Croke06} have applied it, as an example, to a set of three equiprobable symmetric pure states of individual two-dimensional quantum systems, i.e., \emph{qubits}. Later, this case was experimentally demonstrated using qubits encoded into single-photon polarization \cite{Mosley06}. In the present work our goal is to extend this study to individual $D$-dimensional quantum systems (with $D>2$), i.e., \emph{qudits}. The nonorthogonal symmetric states of qudits are known to play an important role in QSD \cite{Ban97,Chefles98-2,Jimenez07} and many other quantum information protocols \cite{Chefles98-1,Yang09,Roa03,Delgado05}. Motivated by this fact, we study the MC strategy applied to a set of $N$ linearly dependent symmetric qudit states, prepared with equal prior probabilities.  Usually, in this problem, an inconclusive outcome is inevitable, and from the conditions established in Ref.~\cite{Croke06} we find the optimal positive operator valued measure (POVM) that maximizes our confidence in identifying each state in the set and minimizes the probability of obtaining inconclusive results. The physical realization of this POVM is completely determined and we show that after an inconclusive outcome, the input states may be mapped into a new set of $N$ equiprobable symmetric states, restricted, however, to a subspace of the original $D$-dimensional Hilbert space. Therefore, by applying the MC measurement again onto this subspace, we can still gain some information about the input states, although with less confidence than before. As we will discuss, this process may be iterated in as many stages as allowed by the input set, until no additional information can be extracted from an inconclusive result. We shall establish the conditions in which this \emph{sequential maximum-confidence measurement} applies and  show that at each stage,  our confidence in identifying the input states is higher than or equal to the one achieved by the optimal ME measurement applied in that stage. Additionally, the more stages we accomplish (within the maximum allowed), the higher will be the probability that this identification was correct. This type of optimized measurement proposed here does not apply for qubits since that after an inconclusive outcome the input states are projected onto a one-dimensional subspace. 

The optimal sequential MC measurement will be illustrated with an explicit example in the simplest possible case where it applies, which is the discrimination among four symmetric qutrit states. In this case, where there is the possibility of performing a two-stage sequential measurement, we propose an experimental procedure which could implement it. Our scheme is based on single photons and linear optics. The symmetric states are encoded into the propagation modes of a single photon, and, at each stage, the optimized measurements are carried out by using the polarization degree of freedom as an ancillary system and a multiport optical interferometer \cite{Reck94,Zukowski97} with photodetectors at each of its outputs. This scheme is feasible with the current technology.

The remainder of the article is organized as follows: In Sec.~\ref{sec:MCM} we review the basic aspects of MC measurements for discriminating nonorthogonal pure states. In Sec.~\ref{sec:symm} this strategy is applied to the problem of discrimination among symmetric qudit states. In Sec.~\ref{sec:SMC} we specify its physical implementation and introduce the concept of sequential maximum-confidence measurements. In Sec.~\ref{sec:Application} we exemplify the sequential MC measurement by applying it to four qutrit states. In addition, we propose an optical network which could experimentally implement it. Finally, a summary of our results and a brief discussion of their potential applications are given in Sec.~\ref{sec:Conc}.

\section{Maximum-confidence quantum measurements}   \label{sec:MCM}

Often, the optimized measurement strategies in the problem of QSD can be treated, mathematically, within the formalism of POVMs \cite{NielsenBook,BarnettBook,PeresBook,HelstromBook}. A POVM is a set of operators $\{\hat{\Pi}_j\}$ which, in order to form a physically realizable measurement, must satisfy the conditions
\begin{equation}    \label{eq:POVM_conditions}
\hat{\Pi}_j\geq 0 \;\;\;\forall\, j, \hspace{5mm} \sum_j\hat{\Pi}_j=\hat{I},
\end{equation}
where $\hat{I}$ is the identity operator on the Hilbert space of the system. Each POVM element $\hat{\Pi}_j$ corresponds to a measurement outcome $\omega_j$ and the probability of obtaining this outcome by measuring a quantum system in the state $\hat{\rho}$ is given by $P(\omega_j|\hat{\rho})={\rm Tr}(\hat{\Pi}_j\hat{\rho})$.

In the MC measurement for QSD, the figure of merit to be optimized is the probability that the prepared state was $\hat{\rho}_j$, given that the outcome of a measurement was $\omega_j$. This conditional probability, $P(\hat{\rho}_j|\omega_j)$, is interpreted as our confidence in taking the outcome $\omega_j$ to indicate the state $\hat{\rho}_j$, and an optimal measurement should maximize it \cite{Croke06}. Using Bayes' rule and the above observations about measurement, this quantity can be written as 
\begin{equation}    \label{eq:confidence}
P(\hat{\rho}_{j}|\omega_{j})=\frac{P(\hat{\rho}_{j})P(\omega_{j}|\hat{\rho}_{j})}{P(\omega_{j})}
=\frac{p_{j}{\rm Tr}(\hat\Pi_{j}\hat\rho_{j})}{{\rm Tr}(\hat\Pi_{j}\hat\rho)}.
\end{equation}
In this expression, $P(\hat{\rho}_{j})=p_j$ is the known preparation probability for the state $\hat{\rho}_j$; $P(\omega_{j}|\hat{\rho}_{j})={\rm Tr}(\hat\Pi_{j}\hat\rho_{j})$, where $\hat{\Pi}_j$ is the POVM element associated with the outcome $\omega_j$ and $P(\omega_j)={\rm Tr}(\hat\Pi_{j}\hat\rho)$ is the total probability of occurrence for the outcome $\omega_j$, where $\hat{\rho}=\sum_{j=0}^{N-1}p_j\hat{\rho}_j$ is the \emph{a priori} density operator for the system.  

As mentioned in the Introduction, the MC measurement has a closed form solution for any set of states. In particular, for a set of pure states, the POVM element that maximizes the conditional probability in Eq.~(\ref{eq:confidence}) is \cite{Croke06}
\begin{equation}    \label{eq:optimal_POVM}
\hat\Pi_{j}=a_j\hat\rho^{-1}\hat\rho_{j}\hat\rho^{-1},
\end{equation}
with the weighting factor $a_j$ given by 
\begin{equation}    \label{eq:a_j}
a_j=\frac{P(\omega_j)}{{\rm Tr}(\hat\rho^{-1}\hat\rho_{j})}=\frac{{\rm Tr}(\hat\Pi_j\hat\rho)}{{\rm Tr}(\hat\rho^{-1}\hat\rho_{j})}.
\end{equation}
The corresponding maximum confidence that the outcome $\omega_j$ identifies the state $\hat\rho_{j}$ becomes
\begin{equation}   \label{eq:MaxConf_prob}
[P(\hat\rho_{j}|\omega_{j})]_{\rm max} = p_{j}{\rm Tr}\left(\hat\rho_{j}\hat\rho^{-1}\right).
\end{equation}

From Eq.~(\ref{eq:confidence}) one can see that the weighting factor $a_j$ of the optimal POVM element (\ref{eq:optimal_POVM}) has no effect on the maximum confidence with which we can identify the state $\hat{\rho}_j$. Therefore, we can construct each optimal $\hat{\Pi}_j$, independently, up to an arbitrary multiplicative factor, but taking into account the first two constraints of Eq.~(\ref{eq:POVM_conditions}). In some occasions, there is no choice of factors $\{a_j\}$ that enables the set of operators $\{\hat{\Pi}_j\}$ to fulfill the completeness condition in Eq.~(\ref{eq:POVM_conditions}). In these cases, an inconclusive result must be added, with the corresponding POVM element $\hat{\Pi}_?=\hat{I}-\sum_{j=0}^{N-1}\hat{\Pi}_j$ subjected to the constraint $\hat{\Pi}_?\geq 0$. An additional criterion of optimality is to choose a set $\{a_j\}$  which minimizes the probability of inconclusive result.

\section{Maximum-confidence discrimination for symmetric states}     \label{sec:symm}

In this section we apply the MC strategy to discriminate among nonorthogonal symmetric pure states of a \emph{single-qudit} system.  Before doing so, let us motivate the problem by defining the symmetric states and their importance in the context of QSD and practical applications of quantum information. A set of $N$ pure states $\{|\psi_j\rangle\}$ spanning a $D$-dimensional Hilbert space, $\mathcal{H}_D$, is called symmetric if there exists a unitary transformation $\hat{U}$ on $\mathcal{H}_D$ such that \cite{Ban97,Chefles98-2}
\begin{subequations}     \label{eq:def_sym}
\begin{eqnarray}
|\psi_j\rangle & = & \hat{U}|\psi_{j-1}\rangle=\hat{U}^j|\psi_{0}\rangle, \label{eq:def_sym_a} \\
|\psi_0\rangle & = & \hat{U}|\psi_{N-1}\rangle, \label{eq:def_sym_b} \\
\hat{U}^N & = & \hat{I}_D. \label{eq:def_sym_c}
\end{eqnarray}
\end{subequations}
If $\{|l\rangle\}$ ($l=0,\cdots,N-1$) is an orthonormal basis in which $\hat{U}$ is diagonal, then from its unitarity and the condition (\ref{eq:def_sym_c}), this operator can be written as
\begin{equation}    \label{eq:unit}
\hat{U}=\sum_{l=0}^{D-1}e^{2\pi il/N}|l\rangle\langle l|.
\end{equation}
Expanding $|\psi_0\rangle$ in the basis $\{|l\rangle\}$ and using Eqs.~(\ref{eq:def_sym}) and (\ref{eq:unit}), the symmetric states will be given by
\begin{equation}   \label{eq:symm_state}
|\psi_0\rangle=\sum_{k=0}^{D-1}c_k|k\rangle \;\rightarrow\; |\psi_j\rangle=\sum_{k=0}^{D-1}c_ke^{2\pi ijk/N}|k\rangle,
\end{equation}
where $\sum_k|c_k|^2=1$ and, without loss of generality, we will assume that all of the $c_k$ are nonzero. These states play a very important role in the development of QSD. In general, for a given discrimination strategy, finding the optimal POVM for an arbitrary set of states is a highly nontrivial task. However, the equiprobable symmetric states provide analytical solutions for many of those strategies, as, for instance, ME \cite{HelstromBook,Ban97} and UD \cite{Chefles98-2,Jimenez07}. In addition the problem of discriminating among symmetric states of qudits naturally arises in some quantum information protocols like entanglement concentration \cite{Chefles98-1,Yang09} and quantum teleportation \cite{Roa03} and entanglement swapping \cite{Delgado05} via nonmaximally entangled channels.  

Let us now apply the MC strategy to a set of $N$ symmetric qudit states, prepared with equal \emph{a priori} probabilities $p_j=1/N$. Here we consider only the case of linearly dependent states ($N>D$) since for linearly independent ones the problem reduces to UD and has already been solved \cite{Chefles98-2}. Using Eq.~(\ref{eq:symm_state}), the \emph{a priori} density operator for this set will be written as
\begin{equation}    \label{eq:rho_prior}
\hat{\rho}=\frac{1}{N}\sum_{j=0}^{N-1}|\psi_{j}\rangle\langle\psi_{j}|= \sum_{k=0}^{D-1}|c_{k}|^{2}|k\rangle\langle k|.
\end{equation}
As we assumed that $c_k\neq 0$ for all $k$, we have $\hat{\rho}^{-1}=\sum_{k=0}^{D-1}|c_{k}|^{-2}|k\rangle\langle k|$, and the maximum confidence calculated from Eq.~(\ref{eq:MaxConf_prob}) will be
\begin{equation}   \label{eq:MaxConf_prob_symm}
[P(\hat\rho_{j}|\omega_{j})]_{\rm max} = \frac{D}{N} \;\;\;\;\;\forall\;j.
\end{equation}
Therefore, our confidence that the symmetric state $\hat\rho_{j}=|\psi_j\rangle\langle\psi_j|$ was indeed present when the outcome $\omega_j$ is obtained will be $D/N$ for each state in the set. The corresponding POVM element that maximizes the confidence is calculated from Eq.~(\ref{eq:optimal_POVM}) to be
\begin{equation}   \label{eq:POVM_elem}
\hat{\Pi}_j = a_j|\phi_j\rangle\langle\phi_j|,
\end{equation}
where $|\phi_j\rangle$ are non-normalized states of the form
\begin{equation}    \label{eq:reciprocal}
|\phi_j\rangle=\sum_{k=0}^{D-1} (c^*_k)^{-1}\hat{U}^j|k\rangle.
\end{equation}
These states are also linearly dependent and symmetric with respect to the transformation $\hat{U}$ given by Eq.~(\ref{eq:unit}). 

It can be shown that the set of operators (\ref{eq:POVM_elem}) will form a POVM only if $|c_j|=1/\sqrt{D}$ and $a_j=1/ND$ for all $j$.  This POVM is the square-root measurement which is the optimal ME measurement \cite{Ban97}. Therefore, in this particular case, the inconclusive outcome will not be necessary and MC and ME strategies coincide. On the other hand, when the magnitude of the state coefficients are not all the same, those operators will not form a POVM for any choice of $\{a_j\}$, and we must include an inconclusive outcome with the POVM element given by $\hat{\Pi}_?=\hat{I}_D-\sum_{j=0}^{N-1}a_j|\phi_j\rangle\langle\phi_j|$. 
For the specific problem under study, this operator can be simplified by noting that the factors $a_j$, given by Eq.~(\ref{eq:a_j}), are proportional to the total probability of occurrence for the outcome $\omega_j$. As the input states (\ref{eq:symm_state}) are symmetric and equally likely, and the measurement states (\ref{eq:reciprocal}) are also symmetric, this total probability $P(\omega_j)$ should be the same for each outcome $\omega_j$. Thus, for some positive constant $a$, we will have $a_j= a$ for all $j$ \cite{Croke06} and the inconclusive POVM element will be written as
\begin{eqnarray}
\hat\Pi_{?} & =  & \hat{I}_D - a\sum_{j=0}^{N-1}|\phi_j\rangle\langle\phi_j| \nonumber \\ &=&\sum_{m=0}^{D-1}\left(1-\frac{aN}{|c_m|^2}\right)|m\rangle\langle m|.
\label{eq:POVM_?}
\end{eqnarray}
The constraint $\hat\Pi_{?}\geq 0$ imposes that $a\leq|c_m|^2/N$ for all $m=0,\cdots,D-1$, and in order to optimize the process we must choose the value of $a$ which minimizes the probability of obtaining an inconclusive result. Using Eqs.~(\ref{eq:rho_prior}) and (\ref{eq:POVM_?}), this probability is calculated to be $P(?) = {\rm Tr}(\hat{\Pi}_?\hat{\rho}) =1-aND$. Its minimum value, subject to $\hat\Pi_{?}\geq 0$, will be achieved when $a=c_{\rm min}^2/N$, where $c_{\rm min}\equiv\min\{|c_j|\}_{j=0}^{D-1}$, and is given by  \begin{equation}    \label{eq:min_?}
[P(?)]_{\rm min} = 1-Dc_{\rm min}^2.
\end{equation}
The corresponding POVM element (\ref{eq:POVM_?}) becomes
\begin{equation}       \label{eq:POVM_?_opt}
\hat\Pi_{?} = \sum_{m=0}^{D-1}\left(1-\frac{c_{\rm min}^2}{|c_m|^2}\right)|m\rangle\langle m|.
\end{equation}

Therefore, we have determined the maximum possible confidence (\ref{eq:MaxConf_prob_symm}) of identifying each state from a linearly dependent set of equiprobable symmetric states and the minimum probability (\ref{eq:min_?}) of obtaining an inconclusive outcome in the process. The corresponding POVM that optimizes this measurement, $\{\hat{\Pi}_j,\hat{\Pi}_?\}$, is given by Eqs.~(\ref{eq:POVM_elem}) and (\ref{eq:POVM_?_opt}). For the case of three symmetric qubit states, the above results reproduce those obtained by Croke \emph{et al.} in Ref.~\cite{Croke06}, as it should be.

\subsection{Maximum confidence in the ME strategy}
\label{subsec:MC_ME}

We can draw a comparison between MC and ME strategies, regarding the confidence achieved in each one for identifying a state as the result of a measurement. For a set of equiprobable symmetric states (\ref{eq:symm_state}) the optimal ME measurement is given by \cite{Ban97}
\begin{equation}     \label{eq:POVM_ME}
\hat{\Pi}^{\rm ME}_j=\frac{D}{N}|\mu_j\rangle\langle\mu_j|,
\end{equation}
where
\begin{equation}     \label{eq:ME_state}
|\mu_j\rangle=\frac{1}{\sqrt{D}}\sum_{k=0}^{D-1} e^{i\arg(c_k)}e^{2\pi ijk/N}|k\rangle.
\end{equation}
The confidence [see Eq.~(\ref{eq:confidence})] that the input state was indeed $\hat{\rho}_j$ when the outcome of this measurement is $\omega_j$ can be calculated with the help of Eqs.~(\ref{eq:symm_state}) and (\ref{eq:rho_prior}), and will be given by 
\begin{equation}   \label{eq:MC_in_ME_general}
[P(\hat\rho_{j}|\omega_{j})]_{\rm max}^{\rm ME} = \frac{1}{N}\left(\sum_{m=0}^{D-1}|c_m|\right)^2 \;\;\;\;\;\forall\;j.
\end{equation}
Employing the Lagrange multiplier method, it is possible to show that $\sum_m|c_m|\leq\sqrt{D}$ and, hence,
\begin{equation}    \label{eq:MC_in_ME}
[P(\hat\rho_{j}|\omega_{j})]_{\rm max}^{\rm ME}  
\leq\frac{D}{N} . 
\end{equation}
Therefore, when the MC measurement is not inconclusive, the confidence that it provides [Eq.~(\ref{eq:MaxConf_prob_symm})] will always be higher than that achieved in the ME measurement. The only exception occurs when $|c_j|=1/\sqrt{D}$ for all $j$. In this case, MC and ME strategies coincide, as we discussed before, and the equality in Eq.~(\ref{eq:MC_in_ME}) holds. In the next section, we will see that the optimized ME measurement (\ref{eq:POVM_ME}) also applies in one step of the optimized MC measurement for discriminating equiprobable symmetric qudit states.

\section{Realization of maximum-confidence measurements for symmetric qudit states}
\label{sec:SMC}

In this section we describe, abstractly, the implementation of the MC measurements for discriminating symmetric qudit states discussed previously. We begin by noting that, as pointed out by Croke \emph{et al.} \cite{Croke06,Croke08}, the MC strategy can be thought of as a two-step process. In the first step, a two-outcome measurement is performed where one outcome is associated with the success ($\omega_{\rm succ}$) and the other with the failure ($\omega_{\rm fail}$) of the process. If $\omega_{\rm succ}$ is obtained, the result is interpreted as successful in the sense that the input states undergo a transformation which enables their identification with maximum confidence by a proper measurement implemented in the  \emph{second} step. On the other hand, if $\omega_{\rm fail}$ is obtained, the result is interpreted as failure (or inconclusive) in the sense that the transformed input states cannot be identified (at all or with maximum confidence, as we will see later) by any further measurement. Therefore, the outcome $\omega_{\rm fail}$ occurs with probability $P(?)$ and is associated with the POVM element $\hat{\Pi}_?$. Accordingly, the outcome $\omega_{\rm succ}$ occurs with probability $P_{\rm s}=1-P(?)$ and is associated with the POVM element $\hat{\Pi}_{\rm s}=\hat{I}_D-\hat{\Pi}_?$.

The whole description of the discrimination process above can be made clearer by resorting to the effect (or detection) operators, $\hat{A}_\ell$ and $\hat{A}^\dagger_\ell$  \cite{BarnettBook}. When a measurement associated with the POVM $\{\hat{\Pi}_\ell\}$ is performed on the state $\hat{\rho}$ and the result is $\ell$, the postmeasurement state changes as $\hat{\rho}\rightarrow\hat{A}_\ell\hat{\rho}\hat{A}_\ell^\dagger/{\rm Tr}(\hat{A}_\ell\hat{\rho}\hat{A}_\ell^\dagger)$. Therefore, the pair of operators $\hat{A}_\ell$ and $\hat{A}^\dagger_\ell$ transform the initial state according to the outcome of a measurement. In terms of these operators, any given POVM element can be written as $\hat{\Pi}_\ell=\hat{A}^\dagger_\ell\hat{A}_\ell$, and the knowledge of $\hat{\Pi}_\ell$ allows us to determine the effect operators, up to an arbitrary unitary transformation $\hat{W}$, through the relation $\hat{A}_\ell=\hat{W}\hat{\Pi}_\ell^{1/2}$. For our particular problem, in the first step of the process, we have to implement a two-outcome POVM given by $\{\hat{\Pi}_{\rm s},\hat{\Pi}_?\}$. The effect operators $\hat{A}_{\rm s}$ and $\hat{A}_?$ associated with the outcomes $\omega_{\rm succ}$ and $\omega_{\rm fail}$, respectively, are obtained from the operator $\hat{\Pi}_?$ [Eq.~(\ref{eq:POVM_?_opt})] and are given by
\begin{equation}    \label{eq:det_op_succ}
\hat A_{\rm s} = \hat{W} \sum_{m=0}^{D-1}\frac{c_{\rm min}}{|c_m|}|m\rangle\langle m|,
\end{equation}
\begin{equation}     \label{eq:det_op_?}
\hat A_{?} = \hat{W}\sum_{m=0}^{D-1}\sqrt{1-\frac{c_{\rm min}^2}{|c_m|^2}}|m\rangle\langle m|.
\end{equation}
Thanks to our freedom in designing the unitary transformation $\hat{W}$, a convenient choice for our purposes will be
\begin{equation}     \label{eq:W_unit}
\hat{W} =\sum_{k=0}^{D-1}e^{-i\arg(c_k)}|k\rangle\langle k|.
\end{equation}
This operator simply removes the relative phases of the postmeasurement states which are associated with the input-state coefficients $\{c_k\}$ (\ref{eq:symm_state}). As we will see, this simplifies the discrimination measurement to be implemented in the second step.

\subsection{Implementation of the two-outcome POVM}
\label{subsec:2_OC_POVM}

The physical implementation of a POVM requires the extension of the Hilbert space of the system to be measured \cite{PeresBook}. This can be provided either by an ancillary quantum system (\emph{ancilla}) or by adding unused extra dimensions of the original system (if they exist). The first method is called tensor product extension (TPE) and the second direct sum extension (DSE) \cite{Chen07}. In either case, the POVM is implemented through a unitary operation acting on the extended space followed by a projective measurement on the ancilla system (TPE) or the entire extended space (DSE). This procedure is based on Neumark's theorem \cite{PeresBook,Neumark40}.

To implement the two-outcome POVM $\{\hat{\Pi}_{\rm s},\hat{\Pi}_?\}$ required in the first step of the MC measurement, we will consider the TPE method. Therefore, we introduce a two-dimensional ancillary system whose Hilbert space is spanned by the logical basis $\{|0\rangle_{\rm a},|1\rangle_{\rm a}\}$. In terms of the effect operators $\hat{A}_{\rm s}$ (\ref{eq:det_op_succ}) and $\hat{A}_?$ (\ref{eq:det_op_?}), the unitary transformation that couples the original $D$-dimensional system and the ancilla can be written as \cite{He06}
\begin{equation}
\hat{\mathcal{U}} = \hat A_{\rm s}\otimes\hat{I}_{\rm a}-\hat A_{?}\otimes i\hat{\sigma}^{\rm a}_y,\label{eq:U}
\end{equation}
where $\hat{I}_{\rm a}$ is the identity and $\hat{\sigma}^{\rm a}_y=-i(|0\rangle\langle 1|-|1\rangle\langle 0|)$ is the Pauli $Y$ operator, both acting on the ancilla space. Without loss of generality, let us assume that the qudit [in the state $|\psi_j\rangle$ of Eq.~(\ref{eq:symm_state})] and the ancilla are, initially, independent and the latter is prepared in the state $|0\rangle_{\rm a}$. Thus, the initial state of the composite system will be $|\psi_j\rangle|0\rangle_{\rm a}$. Applying the unitary transformation of Eq.~(\ref{eq:U}) onto this state and using Eqs.~(\ref{eq:symm_state}), (\ref{eq:min_?}), and (\ref{eq:det_op_succ})--(\ref{eq:W_unit}) we obtain
\begin{eqnarray}
\hat{\mathcal{U}}(|\psi_j\rangle|0\rangle_{\rm a})&=&\hat A_{\rm s}|\psi_j\rangle|0\rangle_{\rm a}+\hat A_{?}|\psi_j\rangle|1\rangle_{\rm a},\nonumber\\[1.5mm]
&=&\sqrt{1-P(?)}|u_j\rangle|0\rangle_{\rm a}+\sqrt{P(?)}|\xi_j\rangle|1\rangle_{\rm a}, \nonumber\\
\label{eq:U2}
\end{eqnarray}
where $P(?)$ is the (minimum) probability of obtaining an inconclusive result given by Eq.~(\ref{eq:min_?}). The qudit states $|u_j\rangle$ and $|\xi_j\rangle$, associated with the transformation of the initial state $|\psi_j\rangle$ by the effect operators $\hat{A}_{\rm s}$ and $\hat{A}_?$, respectively, are given by 
\begin{equation}     \label{eq:u_states}
|u_j\rangle=\frac{1}{\sqrt D}\sum_{m=0}^{D-1}e^{2\pi ijm/N}|m\rangle,
\end{equation}
and
\begin{eqnarray}
|\xi_j\rangle& = & \sum_{m=0}^{D-1}\sqrt{\frac{|c_m|^2-c_{\rm min}^2}{P(?)}}\;e^{2\pi ijm/N}|m\rangle \nonumber\\
& = & \sum_{m=0}^{D-1}C_me^{2\pi ijm/N}|m\rangle,
\label{eq:xi_states}
\end{eqnarray}
with $\sum_{m=0}^{D-1}C_m^2=1$. Both set of states $\{|u_j\rangle\}$ and $\{|\xi_j\rangle\}$ are normalized and nonorthogonal. 

After unitary interaction (\ref{eq:U2}), the POVM  $\{\hat{\Pi}_{\rm s},\hat{\Pi}_?\}$ is accomplished by measuring the ancilla in the logical basis. If it succeeds (fails), i.e., if the outcome $\omega_{\rm succ}$ ($\omega_{\rm fail}$) is obtained, the ancilla and the original system are projected onto $|0\rangle_{\rm a}$ and $|u_j\rangle$ ($|1\rangle_{\rm a}$ and $|\xi_j\rangle$), respectively. From Eq.~(\ref{eq:min_?}) this happens with a success (failure) probability $P_{\rm s}=Dc_{\rm min}^2$ ($P_?=1-Dc_{\rm min}^2$), which is the optimal one. Next, we will discuss how the MC measurement proceeds after obtaining the outcome $\omega_{\rm succ}$ or $\omega_{\rm fail}$ in the realization of the POVM.

\subsection{Success}
\label{subsec_succ}

When the outcome $\omega_{\rm succ}$ is obtained, the input states $|\psi_j\rangle$ [Eq.~(\ref{eq:symm_state})] are mapped into $|u_j\rangle$ [Eq.~(\ref{eq:u_states})]. This occurs with the same probability $[1-P(?)]$ for \emph{any} of the input states.  As we described previously, these transformed states are now subjected to a measurement that will identify each one---and, hence, identify each $|\psi_j\rangle$---with the maximum possible confidence, given by Eq.~(\ref{eq:MaxConf_prob_symm}). This is the second step of the MC measurement. 

The questions that arise now are ``What is the proper measurement?'' and ``How can it be implemented?'' To answer the first, we can recall our discussion in Sec.~\ref{sec:symm} that when the input states are equally likely and the magnitude of their coefficients are $|c_j|=1/\sqrt{D}$ for all $j$, the MC and ME strategies coincide, and so the optimized measurement for both is the same. As the ``input'' states $\{|u_j\rangle\}$ in the second step of the MC strategy satisfy these requirements [see Eq.~(\ref{eq:u_states})], the measurement to distinguish them will be the optimal ME measurement given by Eqs.~(\ref{eq:POVM_ME}) and (\ref{eq:ME_state}). In this case, the POVM element associated with the outcome $j$ will be given by
\begin{equation}   \label{eq:POVM_succ}
\hat{\pi}_j=\frac{D}{N}|u_j\rangle\langle u_j|=|\upsilon_j\rangle\langle\upsilon_j|,
\end{equation}
where $|\upsilon_j\rangle=\sqrt{\frac{D}{N}}|u_j\rangle$ is a non-normalized state. To implement this measurement and, thus, answer the second question, we have to apply Neumark's theorem described above. Here, we will consider the DSE method, which is to say that the original system is, actually, lying on a $D$-dimensional \emph{subspace} of a higher $N$-dimensional Hilbert space. (This approach will be justified in the next section, where we design an optical scheme that implements the MC measurement.) 
Therefore, we look for a unitary operation acting on the entire space that will couple those extra dimensions to the original system and, finally, a projective measurement on this space to accomplish the POVM (\ref{eq:POVM_succ}). To start with, let us consider a set of orthonormal extended states $\{|\upsilon'_j\rangle\}$, defined as
\begin{eqnarray}
|\upsilon'_j\rangle & \equiv & |\upsilon_j\rangle+\frac{1}{\sqrt{N}}\sum_{k=D}^{N-1}e^{2\pi ijk/N}|k\rangle \nonumber\\
&=& \hat{\mathcal{F}}_N|j\rangle, 
\label{eq:extended_v}
\end{eqnarray}
where $\hat{\mathcal{F}}_N$ denotes the discrete Fourier transform acting on the $N$-dimensional space and is given by
\begin{equation}     \label{eq:Fourier}
\hat{\mathcal{F}}_N=\frac{1}{\sqrt N}\sum_{k,l=0}^{N-1}e^{2\pi ikl/N}|k\rangle\langle l|.
\end{equation}
By implementing the POVM $\{\hat{\pi}_j\}$ given by (\ref{eq:POVM_succ}), the probability of obtaining the outcome $j$ if the state was $|u_j\rangle$ will be $P(j|u_j)=\langle u_j|\hat{\pi}_j|u_j\rangle=D/N$. Using  Eq.~(\ref{eq:extended_v}) and writing $\hat{\pi}'_j=|\upsilon'_j\rangle\langle\upsilon'_j|$, it is easy to show that 
\begin{equation}     \label{eq:prob_neumark}
P(j|u_j)=\langle u_j|\hat{\pi}'_j|u_j\rangle=|\langle j|(\hat{\mathcal{F}}_N)^{-1}|u_j\rangle|^2=\frac{D}{N}.
\end{equation}
Therefore, the projective measurement $\{\hat{\pi}'_j\}$ on the extended $N$-dimensional space realizes the POVM $\{\hat{\pi}_j\}$ in the original $D$-dimensional space. From Eq.~(\ref{eq:prob_neumark}) this measurement can be physically understood as follows. First, the inverse Fourier transform is applied to the states $|u_j\rangle$ [Eq.~(\ref{eq:u_states})], yielding  
\begin{equation}    \label{eq:Fourier_state}
(\hat{\mathcal{F}}_N)^{-1}|u_j\rangle =\sqrt{\frac{D}{N}}|j\rangle+\sum_{k=1}^{N-1}\beta_k|j\ominus k\rangle,
\end{equation}
where $\beta_k=\frac{1}{\sqrt{ND}}\sum_{m=0}^{D-1}e^{2\pi imk/N}$ and $\ominus$ denotes subtraction modulo $N$. This transformation provides the unitary coupling among the original system and the additional (and unused) $N-D$ dimensions.  Finally, a projective measurement in the logical basis $\{|j\rangle\}$ (for $j=0,\ldots,N-1$) of the extended space is performed. The probability to find the system in the state  $|j\rangle$, if it was in the transformed state $|u_j\rangle$, is $D/N$. Thus, due to the one-to-one correspondence between $|u_j\rangle$ and $|\psi_j\rangle$, when the outcome $j$ leads us to identify the input state as $|\psi_j\rangle$, our confidence in doing so will be $D/N$, which is the maximum possible  [see Eq.~(\ref{eq:MaxConf_prob_symm})]. This procedure concludes the implementation of the second step of the MC measurement.

\subsection{Failure}
\label{subsec:fail}

When the outcome $\omega_{\rm fail}$ is obtained in the measurement of the POVM $\{\hat{\Pi}_{\rm s},\hat{\Pi}_?\}$, the input states $|\psi_j\rangle$ [Eq.~(\ref{eq:symm_state})] are mapped into $|\xi_j\rangle$ [Eq.~(\ref{eq:xi_states})], which are also symmetric. This outcome is equally likely to occur [with probability $P(?)$] for any of the input states. Thus, the states $|\xi_j\rangle$  also form a set of $N$ equiprobable, symmetric, linearly dependent states. However, they are restricted to a $(D-d)$-dimensional Hilbert space, where $d$ is the multiplicity of the input-state coefficients with minimum magnitude, i.e., $|c_k|=c_{\rm min}$. Since $1\leq d\leq D$, three situations are possible: (i) If $d=D$, the MC and ME strategies coincide, as we discussed before, and in this case there is no inconclusive result [$P(?)=0$ in Eq.~(\ref{eq:min_?})]. (ii) If $d=D-1$, the $N$ states $|\xi_j\rangle$ in Eq.~(\ref{eq:xi_states}) will be identical, up to an irrelevant phase factor. In this case, the failure space is one dimensional and no further measurement will allow us to gain information about the input states. (iii) If $d<D-1$, we are left with a set of $N$ equiprobable symmetric states of $(D-d)$-dimensional qudits. Therefore, we can still gain some information about the input states, if we apply the MC measurement again onto this new set. Doing this, and considering the one-to-one correspondence between the states $|\psi_j\rangle$ and $|\xi_j\rangle$, our maximum confidence in taking an outcome $\omega'_j$ to indicate the input state as $|\psi_j\rangle$ will be given by [see Eq.~(\ref{eq:MaxConf_prob_symm})]
\begin{equation}     \label{eq:MC_subspace}
[P'(\hat{\rho}_j|\omega'_j)]_{\rm max} = \frac{D-d}{N},
\end{equation} 
while the minimum probability of obtaining an inconclusive outcome will be [see Eq.~(\ref{eq:min_?})]
\begin{equation}     \label{eq:?_min_subspace}
[P'(?)]_{\rm min} = 1-(D-d)C_{\rm min}^2,
\end{equation} 
with $\displaystyle C_{\rm min}= \min_{C_j\neq 0} \{C_j\}_{j=0}^{D-1}$ [see Eq.~(\ref{eq:xi_states})]. 

It is important to stress that the maximum confidence (\ref{eq:MC_subspace}) and the minimum failure probability (\ref{eq:?_min_subspace}) refer to the MC measurement performed on the $(D-d)$-dimensional subspace, after an inconclusive outcome has been obtained in the first step of the MC strategy. Obviously, although we can still discriminate the input states, our confidence in doing so will be reduced, otherwise, the first step would not have been optimal. The failure probability, on the other hand, will depend on the input states. 

The implementation of the MC measurement to discriminate between the ``failure'' states $|\xi_j\rangle$---and, hence, discriminate the input states---follows the same procedure described above. For this measurement we have two possibilities. 

\hspace{-2mm} (i)  If the failure states have $C_m=1/\sqrt{D-d}$ for all $m$, there is no inconclusive result, and, hence, the optimized POVM will be the same as the optimal ME POVM [see Eqs.~(\ref{eq:POVM_ME}) and (\ref{eq:ME_state})]. It can be implemented by applying the inverse Fourier transform (\ref{eq:Fourier}) and a projective measurement onto the logical basis $\{|j\rangle\}$, which spans the $N$-dimensional Hilbert space. Doing this, the input states will be identified with the confidence (\ref{eq:MC_subspace}).

\hspace{-2mm} (ii)  If the $C_m$ are not all the same, a two-outcome POVM $\{\hat{\Pi}'_\ell=\hat{A}'^\dagger_\ell\hat{A}'_\ell\}$ $(\ell={\rm s,?})$ is implemented through the unitary coupling (\ref{eq:U}) of the original system with a two-dimensional ancilla, followed by a measurement of the latter. The effect operators $\hat{A}'_{\rm s}$ and $\hat{A}'_?$ will have the same form of those given by Eqs.~(\ref{eq:det_op_succ})--(\ref{eq:W_unit}) but now acting on a $(D-d)$-dimensional subspace and with the coefficients $c_k$ replaced by $C_k$. In case of success, the states $|\xi_j\rangle$ are projected onto states $|u'_j\rangle$, which have the form of Eq.~(\ref{eq:u_states}), but with $D-d$ terms in the sum. Again, due to the one-to-one correspondence between $|u'_j\rangle$ and $|\psi_j\rangle$, the input states will be identified with the confidence of Eq.~(\ref{eq:MC_subspace}) by applying the inverse of the Fourier transform (\ref{eq:Fourier}) to $|u'_j\rangle$ and performing a projective measurement onto the logical basis $\{|j\rangle\}$. On the other hand, in case of failure, the states $|\xi_j\rangle$ are projected onto symmetric states $|\xi'_j\rangle$, which are restricted to a $(D-d-d')$-dimensional subspace, with $d'$ denoting the multiplicity of the failure-state coefficients [see Eq.~(\ref{eq:xi_states})] with minimum magnitude, i.e., $C_m=C_{\rm min}$. The previously described procedure can be applied again over these new failure states and, in accordance with the dimension of its subspace, further MC measurements may or may not gather more information about the input states.

\subsection{Sequential maximum-confidence measurements}
\label{subsec:SMC}

\begin{figure}[t]
\centerline{\includegraphics[width=0.48\textwidth]{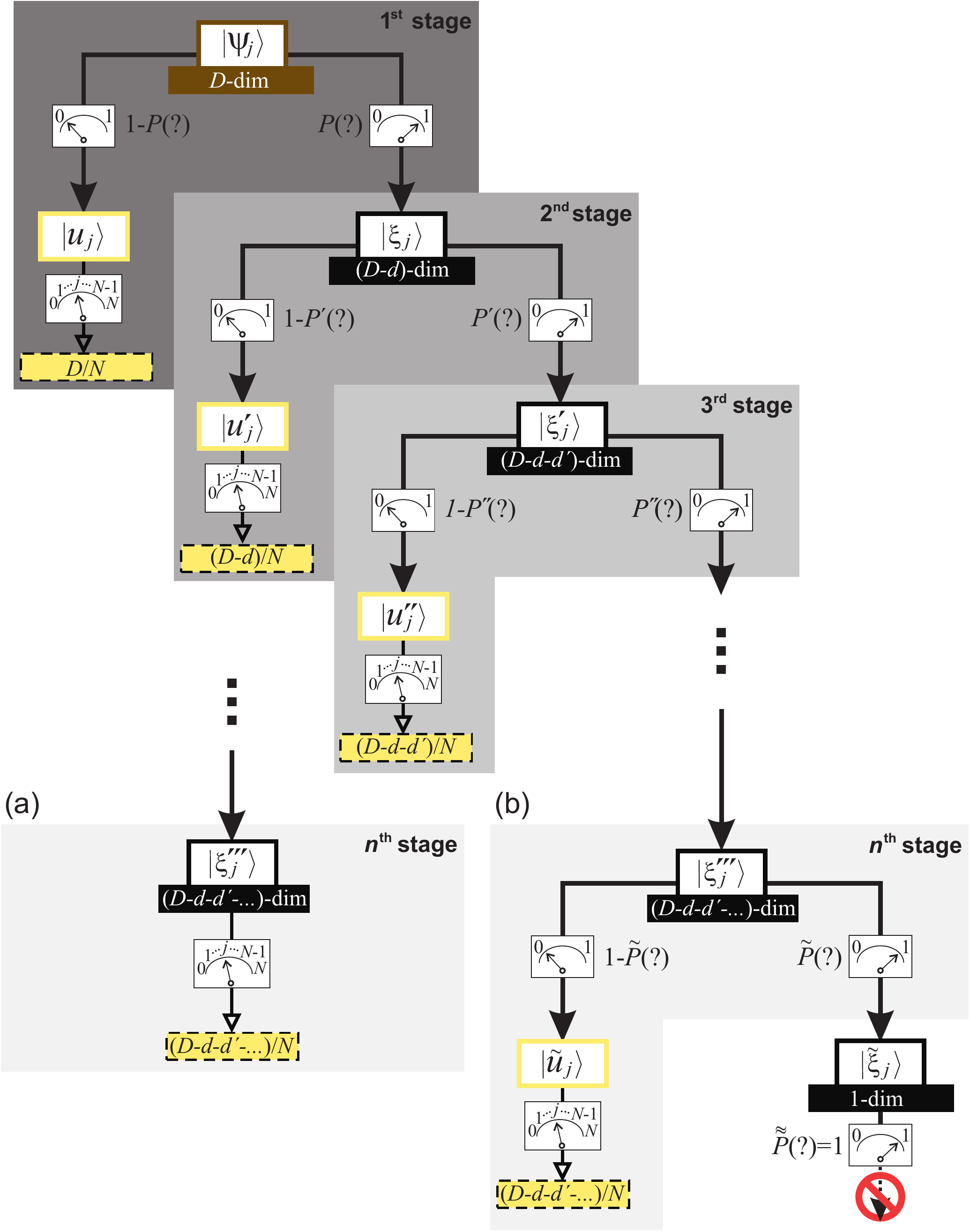}}
\caption{\label{fig:diagram}  (Color online) 
Schematic diagram of an $n$-stage sequential MC measurement for discrimination among $N$ equiprobable symmetric qudit states $\{|\psi_j\rangle\}$. In the first stage, the $D$-dimensional system is coupled to a two-dimensional ancilla [see Eq.~(\ref{eq:U2})] which is measured in the logical basis $\{|0\rangle_{\rm a},|1\rangle_{\rm a}\}$. The outcome ``$0$'', obtained with probability $1-P(?)$, projects the input states onto $\{|u_j\rangle\}$, which are subjected to the $N$-outcome POVM (\ref{eq:POVM_succ}) and correctly identified with probability $D/N$. The outcome ``1'', obtained with probability $P(?)$, projects the input states onto the failure states $\{|\xi_j\rangle\}$, which are restricted to a $(D-d)$-dimensional Hilbert space. The second stage---and each posterior one---starts with the failure states ``$\xi_j$'' of the preceding stage, and the above procedure is repeated within a lower-dimensional space. In each stage, the ``$u_j$'' states are correctly identified with the probability indicated in the dashed box. Due to the one-to-one correspondence between the states ``$u_j$'' and $|\psi_j\rangle$, this probability is the confidence that one identifies $|\psi_j\rangle$ in that stage of the sequential measurement. This sequence ends up in the $n^{\rm th}$ stage when either the magnitude of the failure state coefficients are all equal (a) or the failure space is one dimensional (b). In either case, the total probability of correctly identifying the input states is given by Eq.~(\ref{eq:total_prob}).}
\end{figure}

As we saw above, in case of failure in the first step of the MC measurement, it is possible to implement this discrimination strategy again if the failure space is not one dimensional. This means that, for qudits, there exists the possibility of carrying out \emph{sequential maximum confidence} (SMC) measurements, as illustrated in the schematic diagram of Fig.~\ref{fig:diagram}, for a sequence with $n$ stages. The number of stages in such a measurement ranges from 1 to $D-1$ and will depend on the multiplicities of the input-state coefficients. For instance, if $D=7$ and the magnitude of the input-state coefficients are such that $|c_0|=|c_1|=|c_2|>|c_3|=|c_4|>|c_5|=|c_6|$, it will be possible to implement a three-stage SMC measurement. As can be seen from the diagram of Fig.~\ref{fig:diagram}, at each stage after the first, the ``input'' set is comprised of the failure states from the preceding stage. The two-step MC measurement, as we described earlier, is applied onto these sets at each stage. If it succeeds, the input states $\{|\psi_j\rangle\}$ are discriminated with a given confidence that depends on the dimension of the subspace of the failure states. Otherwise, if it fails, we proceed to next stage. The SMC measurement will end up either when the magnitude of the failure-state coefficients are all equal [Fig.~\ref{fig:diagram}(a)] or the failure space has only one dimension [Fig.~\ref{fig:diagram}(b)]. We note that for qubits, this type of optimized measurement would not be possible since the failure space will always be one dimensional. 

At each stage of the SMC measurement, the ``$u_j$'' states  (see Fig.~\ref{fig:diagram}) are correctly identified with the probability indicated in the dashed boxes. Thanks to the one-to-one correspondence between the ``$u_j$'' and the input states $|\psi_j\rangle$, this probability is exactly our  confidence in taking an outcome $j$ to indicate the input state as $|\psi_j\rangle$, as discussed earlier. For every stage of the SMC measurement we will have that (i) our confidence will be the highest possible in that stage and (ii) it will be greater than the confidence achieved by the optimal ME measurement, if such strategy had been applied in that stage (except
when all the state coefficients have the same magnitude, in which case both confidences are equal).

In addition to the maximum confidence per stage, we can also compute the \emph{total} probability of correctly identifying the input states $|\psi_j\rangle$ after an $n$-stage SMC measurement. Following the diagrams of Fig.~\ref{fig:diagram}, this quantity will be given by 
\begin{eqnarray}
P^{\rm SMC}_{\rm corr}&=&\overbrace{[1-P(?)]\frac{D}{N}}^{1^{\rm st} \;\rm stage}\;+\;\overbrace{P(?)[1-P'(?)]\frac{D-d}{N}}^{2^{\rm nd} \;\rm stage}+\ldots \nonumber\\
&& \text{}+\underbrace{P(?)P'(?)\cdots[1-\tilde{P}(?)]\frac{D-d-d'-\ldots}{N}}_{n^{\rm th} \;\rm stage}, \nonumber\\
\label{eq:total_prob}
\end{eqnarray}
where $\tilde{P}(?)$ is the failure probability at the $n$th stage. The above expression holds for both $n$-stage sequences illustrated in the diagrams of Figs.~\ref{fig:diagram}(a) and \ref{fig:diagram}(b). In the former case, we will have $\tilde{P}(?)=0$ in the $n$th stage. We note that, if one is restricted to implement \emph{only} MC measurements, this probability is already the optimal one since that at each stage the implemented POVM maximizes the confidence and minimizes the probability of inconclusive results. On the other hand, the probability of gaining no information at all about the input states after the $n$ stages will be given by the product of failure probabilities at each stage, that is,
\begin{equation}     \label{eq:prob_???_TOTAL}
P^{\rm SMC}_?=\overbrace{P(?)P'(?)\cdots \tilde{P}(?)}^{n\;\rm stages}.
\end{equation}
If the sequence ends up as sketched in the diagram of Fig.~\ref{fig:diagram}(a) we will have $P^{\rm SMC}_?=0$.

It is clear from Eq.~(\ref{eq:total_prob}) that, although our confidence in determining the input states decreases from one stage to the next, our chance of correctly doing so increases the more stages we accomplish within the maximum allowed by the input set. The upper bound of $P^{\rm SMC}_{\rm corr}$ is $D/N$ and it is attained when, in the first stage, there is no inconclusive outcome, that is, $P(?)=0$. This, as already discussed, occurs when the MC and ME strategies coincide. In fact, proceeding with the comparison between these two strategies started in Sec.~\ref{subsec:MC_ME}, we have that 
\begin{equation}     \label{eq:SMC_ME_probs}
P^{\rm SMC}_{\rm corr}\leq P^{\rm ME}_{\rm corr}=[P(\hat{\rho}_j|\omega_j)]^{\rm ME}_{\rm max},
\end{equation}
where $P^{\rm ME}_{\rm corr}=\sum_{j=0}^{N-1}p_j{\rm Tr}(\hat{\rho}_j\hat{\Pi}_j^{\rm ME})$, for $p_j=1/N$ and $\hat{\Pi}_j^{\rm ME}$ given by Eq.~(\ref{eq:POVM_ME}), and $[P(\hat{\rho}_j|\omega_j)]^{\rm ME}_{\rm max}$ is given by Eq.~(\ref{eq:MC_in_ME_general}). Therefore, while the MC measurement gives higher confidences in identifying the input states, the optimal ME measurement gives, on the average, a larger number of correct identifications. Which strategy is more appropriate to apply will depend on the practical situation.

\section{Application to four symmetric qutrit states and an optical implementation}
\label{sec:Application}

The simplest case where the optimal SMC measurement has the possibility of being carried out is that for the discrimination among four symmetric qutrit states. In this section we apply the results obtained previously to analyze this particular problem and, additionally, we propose an optical network which could realize this measurement.

\subsection{SMC measurements for the discrimination of four symmetric qutrit states} 

From the definition given by Eq.~(\ref{eq:symm_state}), a set of $N=4$ symmetric states in a three-dimensional Hilbert space (qutrit) can be written as
\begin{equation}     \label{eq:symm_qutrits}
|\psi_j\rangle=c_0|0\rangle + c_1e^{i\pi j/2}|1\rangle + c_2e^{i\pi j}|2\rangle,
\end{equation}
with $j=0,\ldots,3$, and $\sum_{k=0}^{2}|c_k|^2=1$. First, one implements the optimal two-outcome POVM $\{\hat{A}^{\dagger}_{\rm s}\hat{A}_{\rm s},\hat{A}^{\dagger}_{?}\hat{A}_{?}\}$ as described in Sec.~\ref{subsec:2_OC_POVM}, with $\hat{A}_{\rm s}$ and $\hat{A}_{?}$ given by Eqs.~(\ref{eq:det_op_succ}) and (\ref{eq:det_op_?}), respectively. In case of success, the states $|\psi_j\rangle$ are projected onto
\begin{equation}
|u_j\rangle=\frac{1}{\sqrt{3}}(|0\rangle + e^{i\pi j/2}|1\rangle + e^{i\pi j}|2\rangle).
\label{eq:u_qutrit}
\end{equation}
A four-outcome POVM [see Eq.~(\ref{eq:POVM_succ})] then is implemented with these states, and, for each possible outcome $\omega_j$, we have the same maximum confidence that the input state was $|\psi_j\rangle$. From Eq.~(\ref{eq:MaxConf_prob_symm}), it will be given by
\begin{equation}    \label{eq:MC_3_4}
[P(\hat{\rho}_j|\omega_j)]_{\rm max}=\frac{3}{4}.
\end{equation}
In case of failure, the states $|\psi_j\rangle$ are projected onto  
\begin{equation}     \label{eq:failure_qutrit} 
|\xi_j\rangle = C_0|0\rangle + C_1e^{i\pi j/2}|1\rangle + C_2e^{i\pi j}|2\rangle,
\end{equation}
where $C_j=\sqrt{\frac{|c_j|^2-c_{\rm min}^2}{P(?)}}$ for all $j$. We note that at least one of the coefficients $C_j$ vanishes. From Eq.~(\ref{eq:min_?}), the minimum failure probability will be 
\begin{equation}     \label{eq:Prob_?_qutrit}
P(?)= 1-3[\min(|c_0|,|c_1|,\sqrt{1-|c_0|^2-|c_1|^2})]^2.
\end{equation}

A second stage of MC measurement will be applicable only if the multiplicity of the input-state coefficients $\{c_j\}$ with minimum magnitude is one. When this is the case, the failure states $|\xi_j\rangle$ in Eq.~(\ref{eq:failure_qutrit}) will form a set of four nonorthogonal symmetric qubit states. Once again, the two-step MC measurement is applied to this new set and, if it succeeds, we can identify the input states with the confidence
\begin{equation}    \label{eq:MC_1_2}
[P'(\hat{\rho}_j|\omega'_j)]_{\rm max}=\frac{1}{2}.
\end{equation}
Otherwise, if it fails, with the minimum probability
\begin{equation}     \label{eq:Prob_?_qutrit_2}
P'(?)= 1-2\left[\min_{C_j\neq 0}(C_0,C_1,C_2)\right]^2,
\end{equation}
there is no chance of gaining more information about the input states, since the failure states will be projected onto a one-dimensional subspace.

From Eqs.~(\ref{eq:total_prob}) and (\ref{eq:prob_???_TOTAL}), the total probability of correctly identifying the input states, and the probability of obtaining no information at all about them will be given, respectively, by
\begin{equation}    \label{eq:prob_SMC_qutrit}
P^{\rm SMC}_{\rm corr}=\overbrace{[1-P(?)]\frac{3}{4}}^{1^{\rm st}\;\rm stage}\ + \ \overbrace{P(?)[1-P'(?)]\frac{1}{2}}^{2^{\rm nd}\;\rm stage},
\end{equation}
\begin{equation}    \label{eq:prob_???_qutrit}
P^{\rm SMC}_?=\overbrace{P(?)P'(?)}^{2\;\rm stages},
\end{equation}
with $P(?)$ and $P'(?)$ given by Eqs.~(\ref{eq:Prob_?_qutrit}) and (\ref{eq:Prob_?_qutrit_2}), respectively. For completeness, one can also compute the probability of making an erroneous identification, which is $P_{\rm err}^{\rm SMC}=1-P_{\rm corr}^{\rm SMC}-P_{?}^{\rm SMC}$.

\begin{figure}
\centerline{{\includegraphics[width=0.445\textwidth]{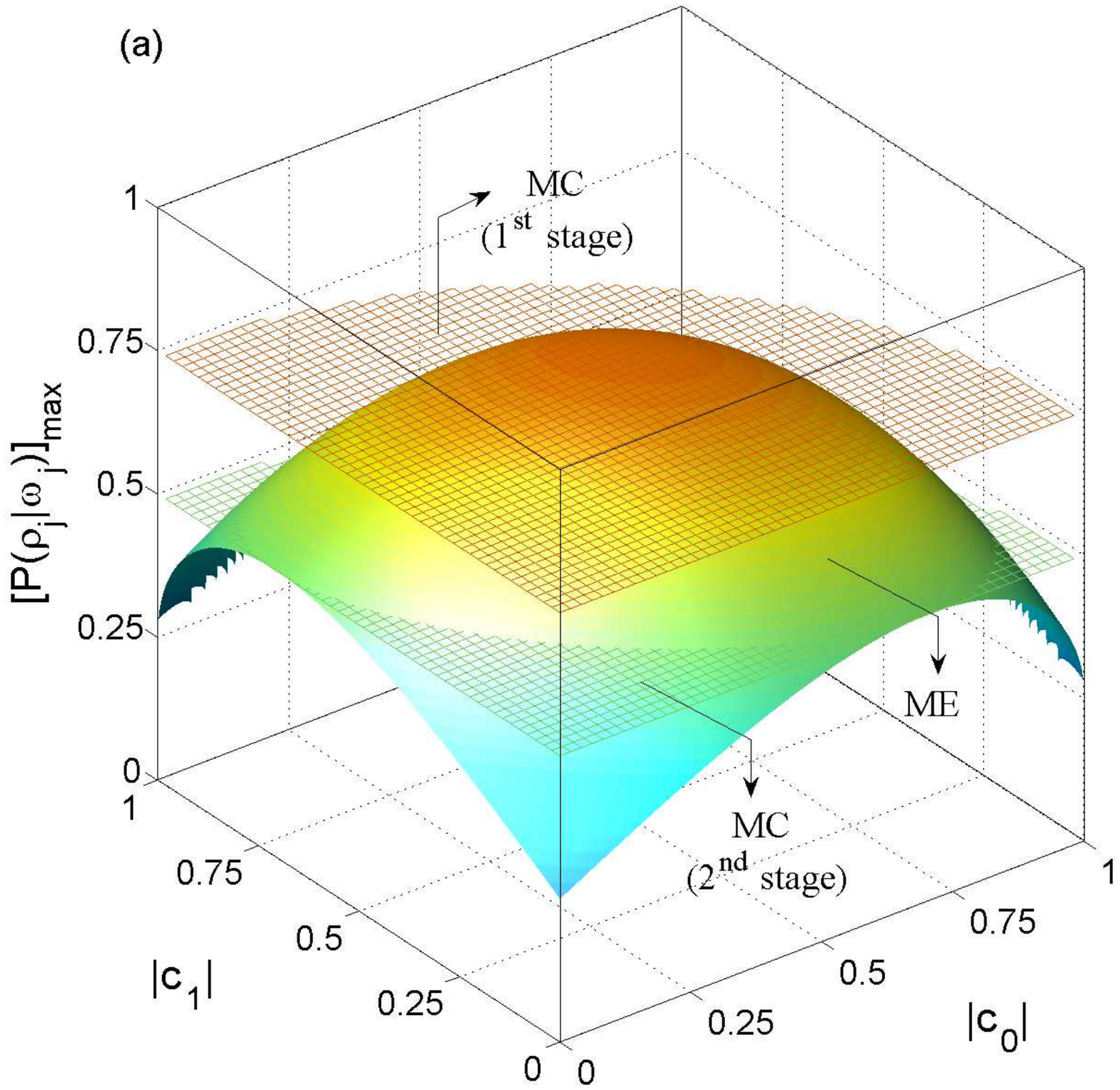}}}
\vspace{2mm}
\centerline{{\includegraphics[width=0.425\textwidth]{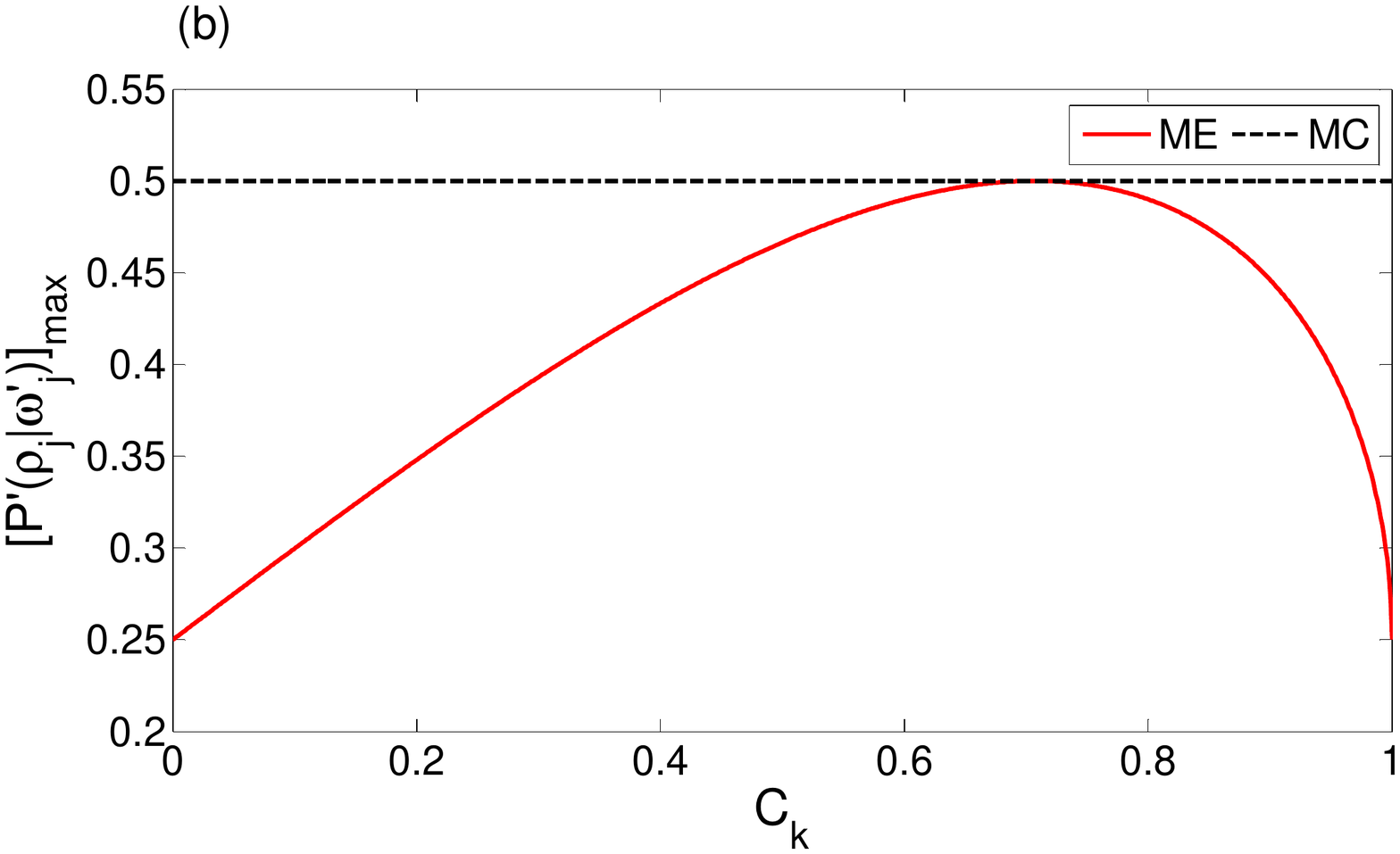}}}
\vspace{-1.9cm}
\caption{  (Color online) (a) Comparison of the maximum-confidence figure of merit for the optimal MC strategy (applied in the first and second stages of the SMC measurement), and the optimal ME strategy, as a function of the magnitude of the input state coefficients (\ref{eq:symm_qutrits}). (b) Comparison of the maximum-confidence figure of merit for the optimal MC strategy applied in the second stage and the optimal ME strategy, if the latter had been applied in the second stage of the SMC measurement. These probabilities are plotted as a function of the magnitude of  either nonvanishing failure state coefficient (\ref{eq:failure_qutrit}). }
	\label{fig:MC_and_ME}
\end{figure}

Let us now analyze graphically the above results. First, we compare the confidence achieved in the identification of the input state as the result of a measurement for ME and MC strategies. The former is calculated from Eq.~(\ref{eq:MC_in_ME_general}) and the latter is given by Eqs.~(\ref{eq:MC_3_4}) (first stage) and (\ref{eq:MC_1_2}) (second stage). Figure~\ref{fig:MC_and_ME}(a) shows the confidence of the optimal ME measurement applied on the input states $|\psi_j\rangle$ in Eq.~(\ref{eq:symm_qutrits}) as a function of the magnitude of their coefficients and also the confidences of the MC measurement applied in each stage. As discussed before, the MC measurement in the first stage always gives a greater confidence than that found for ME (except when the magnitude of the state coefficients are all equal). Interestingly, although our confidence in the second stage becomes smaller, it is still larger than that of ME for many possible sets of input states, as can be seen in Fig.~\ref{fig:MC_and_ME}(a). Figure~\ref{fig:MC_and_ME}(b) shows the confidence in the second stage of the SMC measurement compared with the optimal ME, if the latter had been implemented in that stage. These probabilities were plotted as a function of the magnitude of either nonvanishing failure-state coefficient in Eq.~(\ref{eq:failure_qutrit}). As expected, the MC measurement gives us greater confidence than the ME measurement, except for $C_j=1/\sqrt{2}$.

\begin{figure*}
\centerline{\includegraphics[width=.45\textwidth]{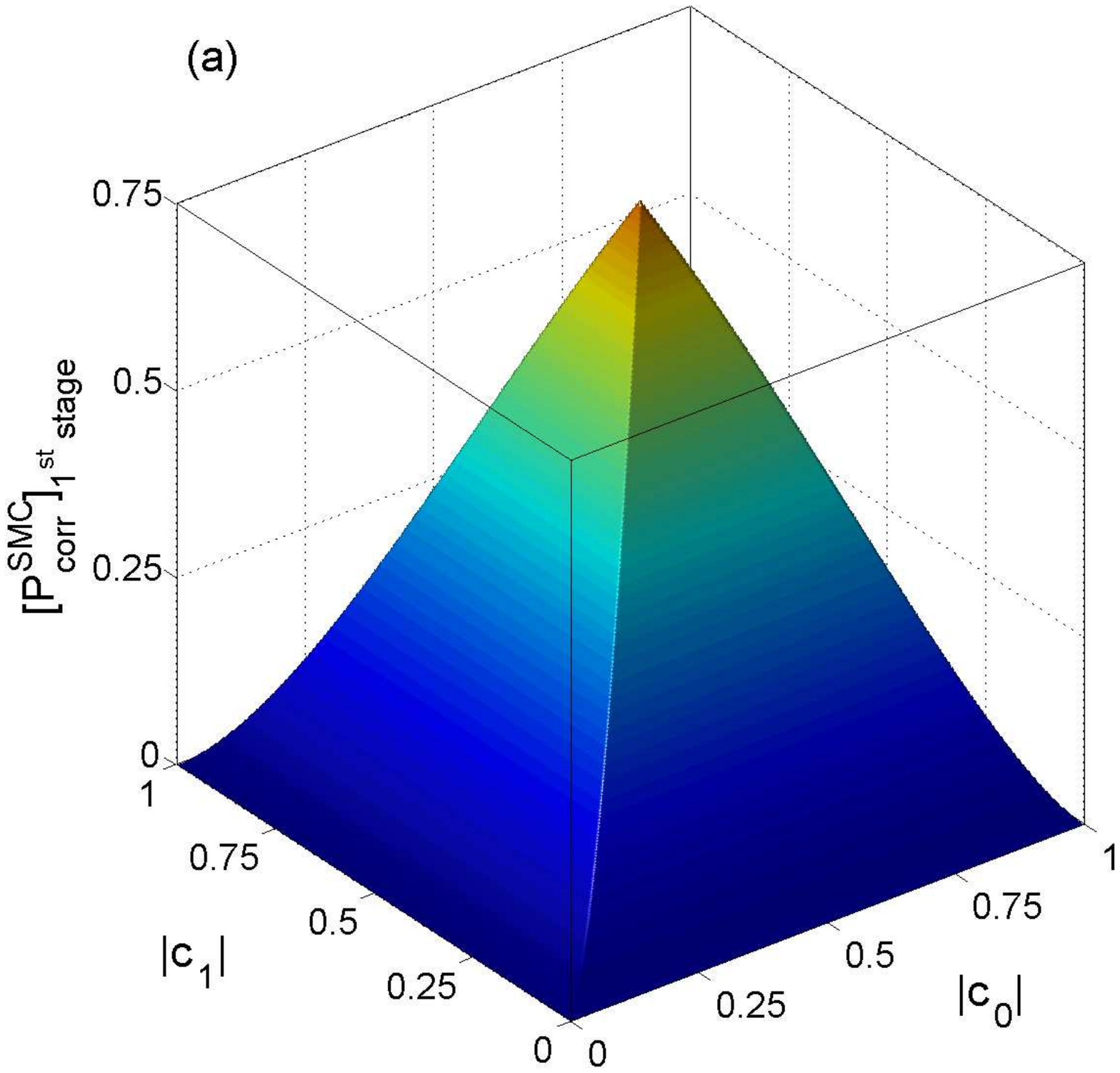}
\includegraphics[width=.45\textwidth]{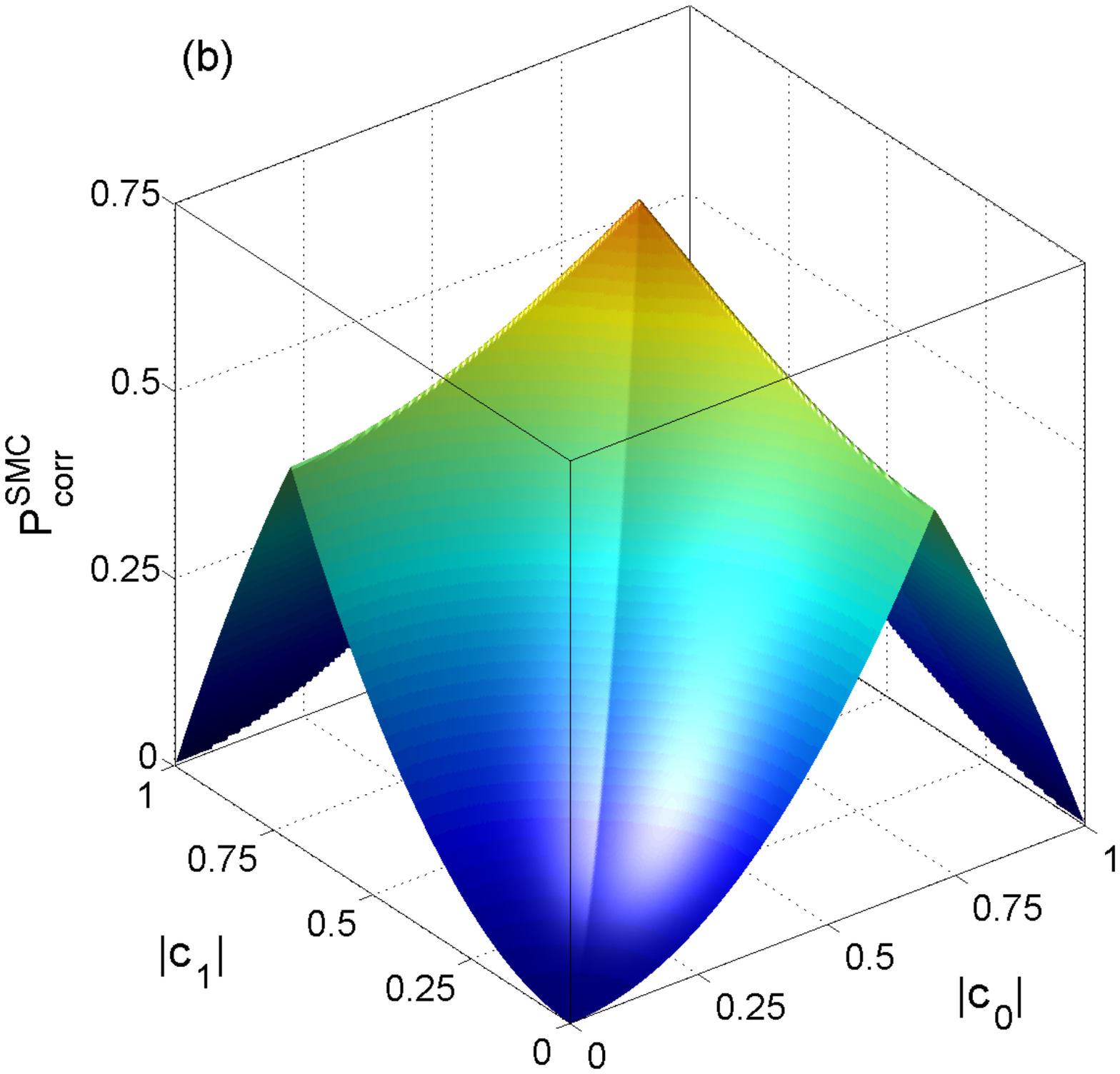}}
\vspace{-2.75cm}
\centerline{\includegraphics[width=.45\textwidth]{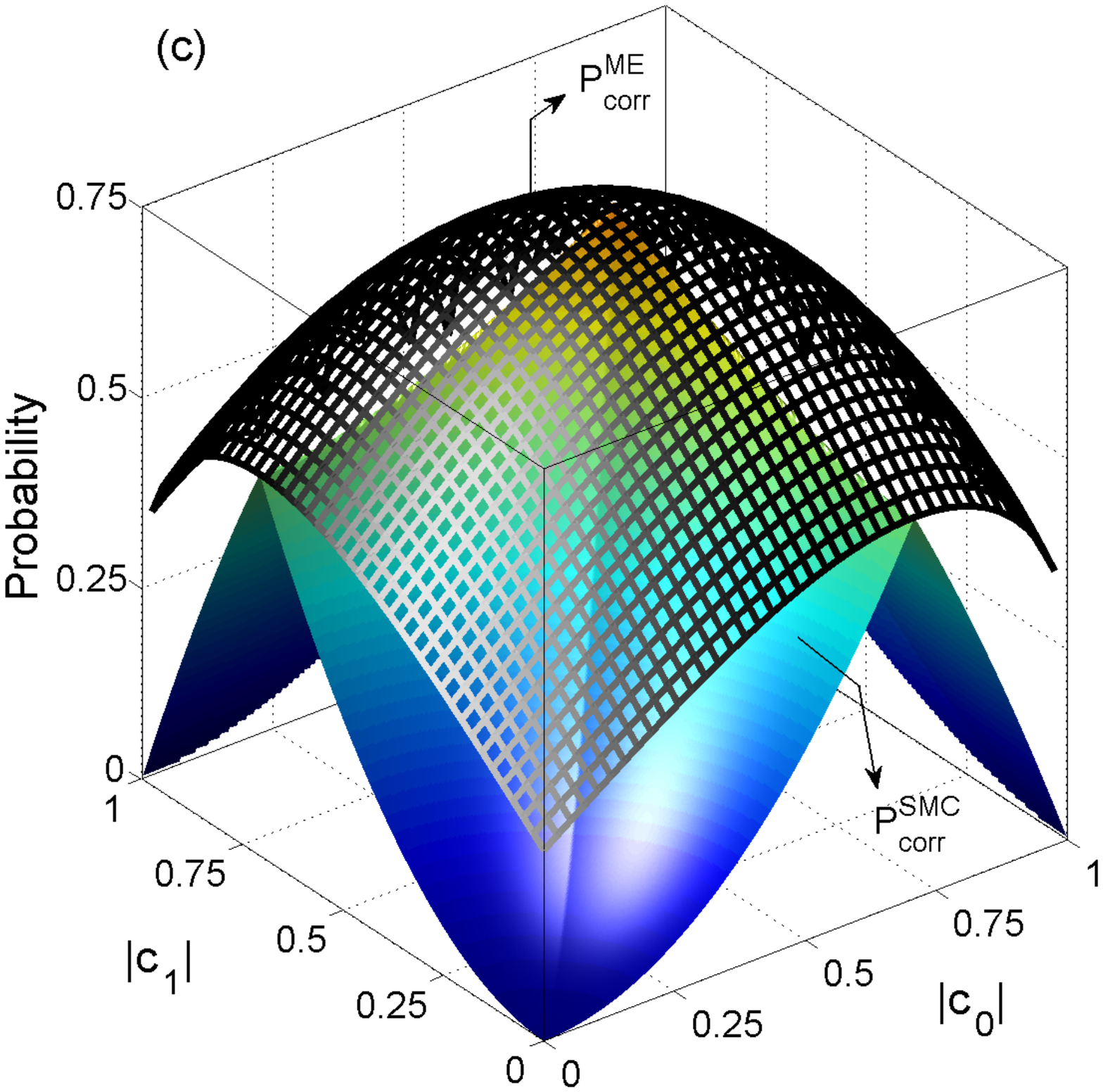}
\includegraphics[width=.45\textwidth]{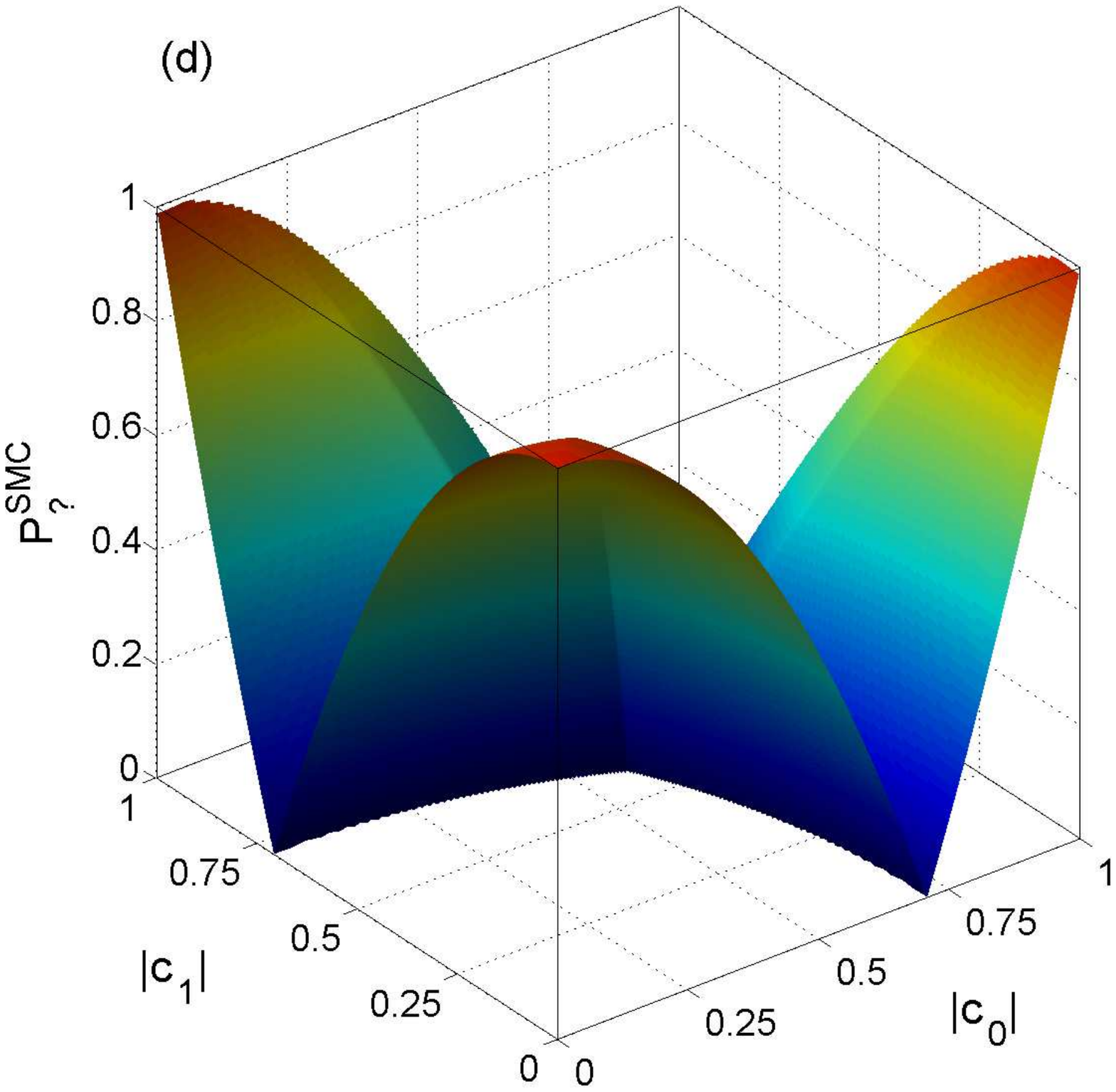}}
\vspace{-2.75cm}
\caption{\label{fig:Prob_corr}  (Color online) Probability of correctly identifying the input states for the first stage (a) and the complete SMC measurement (b). (c) Comparison between the probabilities achieved by the optimal ME and SMC measurements. (d) Probability of gaining no information about the input states after a two-stage SMC measurement. All the probabilities are plotted as a function of the magnitude of the input-state coefficients (\ref{eq:symm_qutrits}). 
 }
\end{figure*}

In the graphics of Figs.~\ref{fig:Prob_corr}(a)--\ref{fig:Prob_corr}(c) we plot the probabilities of correctly identifying the input states, as a function of the magnitude of the input-state coefficients [Eq.~(\ref{eq:symm_qutrits})], achieved in the SMC measurement and in the optimal ME measurement. First, we consider the MC measurement applied only in the first stage, which means that after an inconclusive outcome we make no further measurement. Using Eq.~(\ref{eq:Prob_?_qutrit}) and the first term of the right-rand side of Eq.~(\ref{eq:prob_SMC_qutrit}) we obtain the plot shown in  Fig.~\ref{fig:Prob_corr}(a). Now, taking into account the two stages, the probability $P_{\rm corr}^{\rm SMC}$ in Eq.~(\ref{eq:prob_SMC_qutrit}) is plotted in Fig.~\ref{fig:Prob_corr}(b). By comparing the graphics of Figs.~\ref{fig:Prob_corr}(a) and \ref{fig:Prob_corr}(b), it can be clearly seen that the addition of a second stage significantly increases the chances of gaining information about the input states. For comparison purposes, in Fig.~\ref{fig:Prob_corr}(c) we plot $P_{\rm corr}^{\rm SMC}$ together with $P_{\rm corr}^{\rm ME}$, which is given by Eq.~(\ref{eq:MC_in_ME_general}) with $D=3$. As discussed earlier, the optimal ME measurement is, in general, better when we consider this figure of merit for the discrimination protocol, and the graphics of Fig.~\ref{fig:Prob_corr}(c) corroborate the inequality shown in Eq.~(\ref{eq:SMC_ME_probs}). Finally, using Eqs.~(\ref{eq:Prob_?_qutrit}), (\ref{eq:Prob_?_qutrit_2}), and (\ref{eq:prob_???_qutrit}), in the graphic of Fig.~\ref{fig:Prob_corr}(d) we plot the probability of gaining no information about the input states after a two-stage SMC measurement.

\subsection{Proposed optical network for implementing a two-stage SMC measurement}

We now propose an experimental procedure to implement a two-stage SMC measurement for the discrimination among four equiprobable symmetric qutrit states, discussed above. Our scheme is based on a recent proposal by Jim\'enez \emph{et al.} \cite{Jimenez07} for the experimental realization of UD and ME discrimination among linearly \text{independent} symmetric qudit states. It makes use of single photons to encode the input states $|\psi_j\rangle$, and linear optical elements and photodetectors to carry out the proper transformations and measurements. 

\begin{figure}
\centerline{\includegraphics[width=0.48\textwidth]{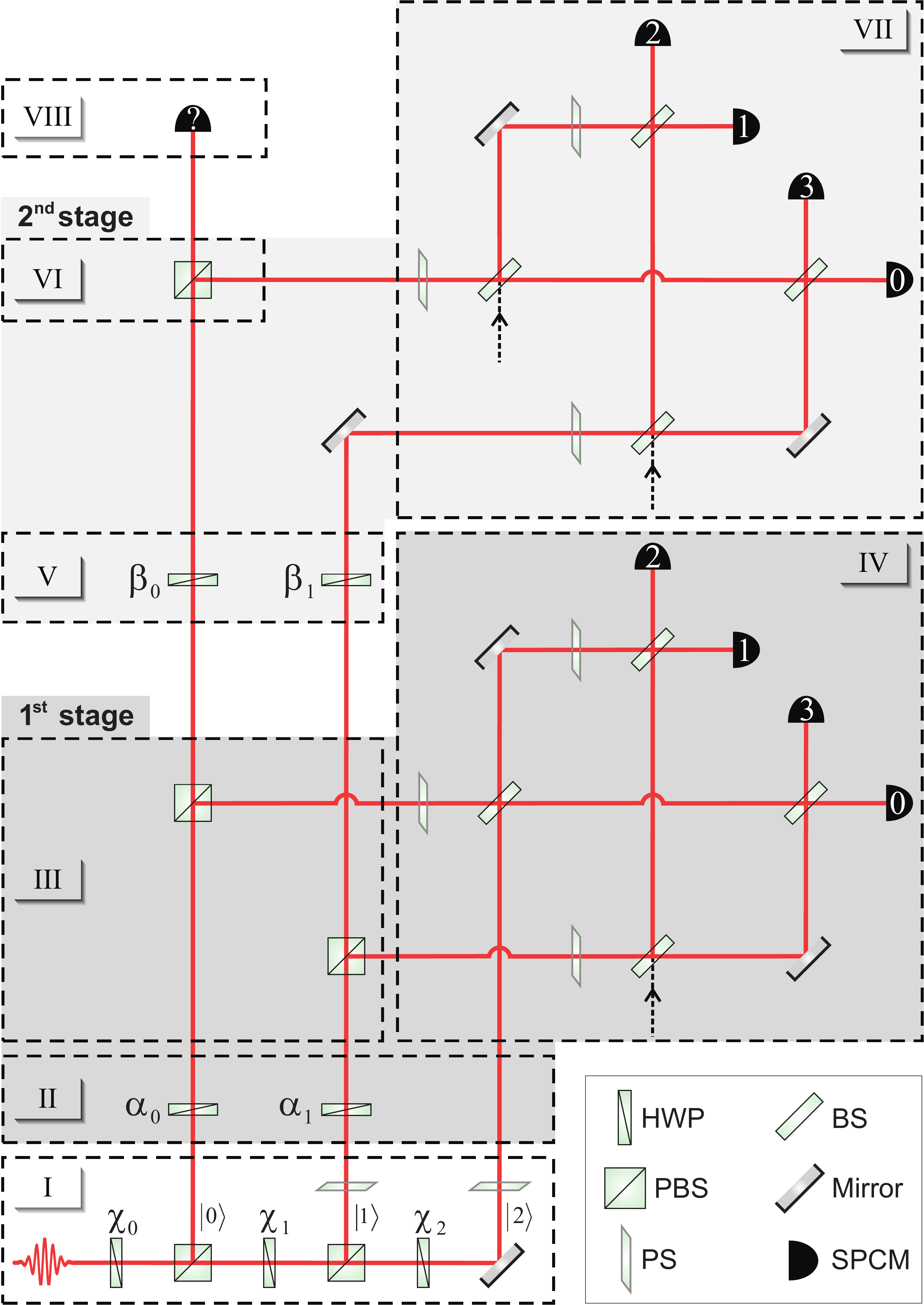}}
\caption{\label{fig:setup}  (Color online) Proposed optical network that implements the optimal two-stage SMC measurement for the discrimination among four symmetric qutrit states. Each dashed box from I to VIII corresponds to a step in the process---from state preparation to measurement---which is discussed in detail in the text. The dark (light) gray shaded region represents the first (second) stage of the SMC measurement. The inset shows the symbols for the single photon counting modules (SPCM) as well as the optical elements used in the scheme. Abbreviations: HWP, half-wave plate; PBS, polarizing beam splitter; PS, phase-shifter; BS, 50:50 beam splitter; mirror.    }
\end{figure}

The sketch of our proposed optical network is shown in Fig.~\ref{fig:setup}. The dashed boxes, numbered from I to VIII, indicate each step of the protocol from state preparation to measurement, while the dark and light gray shaded regions indicate the first and second stages of the SMC measurement, respectively. The qutrit states will be encoded in the propagation modes $\{|0\rangle,|1\rangle,|2\rangle\}$ of a single photon, which could be generated, for example, using either heralding photon-pair sources or on-demand single-photon sources. In this scheme, the orientations of the half-wave plates (HWPs) are specified by the angle the fast axis makes with the horizontal, and all the polarizing beam splitters (PBSs) transmit vertical polarization ($|V\rangle$) and reflect horizontally polarized light ($|H\rangle$); the nonpolarizing beam splitters (BSs) are 50:50, and the phase-shifters (PSs) are adjusted to add the proper phases in the preparation and measurement of the qutrit states. Finally, the detectors at each output are single-photon counting modules (SPCM).

Box I in Fig.~\ref{fig:setup} illustrates the preparation of the input states (\ref{eq:symm_qutrits}). Here, the polarization will be used to assist this preparation. The three half-wave plates (HWPs) are oriented at $\chi_0=\theta/2$, $\chi_1=(\varphi/2+\pi/4$), and $\chi_2=\pi/4$, respectively. Let us assume that the photon is, initially, horizontally polarized. Its quantum state after box I will be $|\psi_j\rangle|H\rangle_{\rm a}$, where $|\psi_j\rangle$ is given by Eq.~(\ref{eq:symm_qutrits}), with $c_0=\cos\theta$, $c_1=\sin\theta\cos\varphi$, and $c_2=\sin\theta\sin\varphi$. For simplicity, and without loss of generality, we set $0\leq\varphi\leq\pi/4$ and $0\leq\theta\leq\tan^{-1}(\sec\varphi)$. Hence, the real-valued input-state coefficients $c_j$ will be ordered as $c_2\leq c_1\leq c_0$, which means that $c_2=c_{\rm min}$.

Having prepared the input state we now proceed to perform the first stage of the SMC measurement in order to identify this state with maximum confidence. As discussed in Sec.~\ref{sec:SMC}, a MC  measurement is a two-step process. The first consists of implementing a two-outcome POVM, which in our scheme is accomplished within the boxes II and III (Fig.~\ref{fig:setup}). Here, the photon polarization will play the role of the required two-dimensional ancilla which will provide the TPE of the qutrit Hilbert space. To implement the optimal unitary transformation (\ref{eq:U}) that couples the original system and ancilla, we make use of HWPs in the modes 0 and 1 (box II) oriented at $\alpha_j=\frac{1}{2}\cos^{-1}(c_2/c_j)$, for $j=0,1$. Upon the identification $|H\rangle_{\rm a}\equiv|0\rangle_{\rm a}$ and $|V\rangle_{\rm a}\equiv|1\rangle_{\rm a}$, the system-ancilla state, after conditional polarization rotations in box II, will be transformed as 
\begin{equation}    \label{eq:qutrit_transf_exp}
|\psi_j\rangle|0\rangle_{\rm a} \;\rightarrow\; \sqrt{3c_2^2}|u_j\rangle|0\rangle_{\rm a} + \sqrt{1-3c_2^2}|\xi_j\rangle|1\rangle_{\rm a},
\end{equation}
where $1-3c_2^2=[P(?)]_{\rm min}$, $|u_j\rangle$ is given by Eq.~(\ref{eq:u_qutrit}) and 
\begin{equation}    \label{eq:xi_exp}
|\xi_j\rangle=\sqrt{\frac{c_0^2-c_2^2}{1-3c_2^2}}|0\rangle + e^{i\pi j/2}\sqrt{\frac{c_1^2-c_2^2}{1-3c_2^2}}|1\rangle.
\end{equation}
To accomplish the POVM the photon polarization is measured in the $\{H,V\}$ basis with PBSs in the modes 0 and 1 (box III). If successful, that is, if the polarization is projected onto $|0\rangle_{\rm a}$, one proceeds to the second step, which is the final measurement that identifies the transformed states $|u_j\rangle$ (and hence $|\psi_j\rangle$) with maximum confidence. According with the discussion in Sec.~\ref{subsec_succ}, in the present example this measurement is a four-outcome POVM implemented by first applying an inverse Fourier transform ($\mathcal{\hat{F}}_4^{-1}$) acting on a four-dimensional Hilbert space [see Eq.~(\ref{eq:Fourier_state})], followed by a projective measurement in the logical basis of this space $\{|0\rangle,|1\rangle,|2\rangle,|3\rangle\}$. The operation $\mathcal{\hat{F}}_4^{-1}$ can be implemented by a symmetric eight-port interferometer \cite{Reck94,Zukowski97,Jimenez07} as the one illustrated in box IV of Fig.~\ref{fig:setup} \cite{CommentFourier}. The vacuum input to this interferometer, indicated by the dashed arrow, provides the DSE of the Hilbert space and allows us to implement the four-outcome POVM, which is finally accomplished by detecting the photon at one of the output ports labeled from 0 to 3. The probability that the photodetector in port $k$ clicks if the input state to the interferometer was $|u_j\rangle$ will be
\begin{equation}
P(k|u_j) =|\langle k|\mathcal{\hat{F}}_4^{-1}|u_j\rangle|^2 = \frac{3}{4}\delta_{jk} + \frac{1}{12}\sum^3_{\stackrel{\scriptstyle l=0}{l\neq j}} \delta_{kl}.
\end{equation}
Therefore, due to the one-to-one correspondence between the states $|u_j\rangle$ and $|\psi_j\rangle$, if a click in the detector $j$ leads us to identify the input state as $|\psi_j\rangle$, we can see from the above equation that our confidence in doing so will be $3/4$, which is the maximum one in this first stage of the SMC measurement  [see Eq.~(\ref{eq:MC_3_4})]. 

When an inconclusive result is obtained from the two-outcome POVM above, the photon polarization in step III is projected onto $|1\rangle_{\rm a}$. Accordingly, one can see from Eq.~(\ref{eq:qutrit_transf_exp}) that the input states are mapped into $\{|\xi_j\rangle\}$ given by Eq.~(\ref{eq:xi_exp}). This new set of $N=4$ equiprobable symmetric qubit states forms the input set for the second stage of the SMC measurement, as illustrated in the light gray shaded region of Fig.~\ref{fig:setup}. Again, a two-outcome POVM (boxes V and VI) is implemented using the polarization as an ancilla. The HWPs in the modes 0 and 1 (box V), oriented at $\beta_{0}=\left[\frac{1}{2}\cos^{-1}\left(\sqrt{\frac{c_{1}^{2} - c_{2}^{2}}{c_{0}^{2} - c_{2}^{2}}}\right)+\frac{\pi}{4}\right]$ and $\beta_1=\pi/4$, provide the optimal unitary coupling (\ref{eq:U}) and, hence, transform the system-ancilla state as
\begin{equation}    \label{eq:xi_transf_exp}
|\xi_j\rangle|1\rangle_{\rm a} \rightarrow\sqrt{\frac{2(c_1^2-c_2^2)}{1-3c_2^2}}|u'_j\rangle|0\rangle_{\rm a} + \sqrt{\frac{c_0^2-c_1^2}{1-3c_2^2}}|0\rangle|1\rangle_{\rm a},
\end{equation}
where $(c_0^2-c_1^2)/(1-3c_2^2)=[P'(?)]_{\rm min}$ and $|u'_j\rangle=\frac{1}{\sqrt{2}}(|0\rangle+e^{i\pi j/2}|1\rangle$. The ancilla then is measured in the $\{H,V\}$ basis with a PBS in the mode 0 (box VI). If it succeeds, the state $|\xi_j\rangle$ is projected onto $|u'_j\rangle$ which will be determined, with the maximum possible confidence in this second stage of the SMC measurement. The four-outcome POVM that identifies $|u'_j\rangle$ (and hence $|\psi_j\rangle$) is implemented by a symmetric eight-port interferometer followed by photodetectors at each of its outputs, as sketched in box VII of Fig.~\ref{fig:setup} \cite{CommentFourier}. Now, in order to implement this POVM, \emph{two} vacuum inputs to the interferometer (indicated by the dashed arrows in box VII) are needed for providing the two extra dimensions via DSE of the Hilbert space. Similarly to the first stage, the probability that the photodetector in port $k$ clicks if the input state to the interferometer was $|u'_j\rangle$ will be
\begin{eqnarray}
P'(k|u'_j) & = & |\langle k|\mathcal{\hat{F}}_4^{-1}|u'_j\rangle|^2 = \frac{1}{2}\delta_{jk} + \frac{1}{4}\sum^3_{l=0} |\epsilon_{j,j\oplus 2,l}|\delta_{kl} , \nonumber\\
&&
\end{eqnarray}
where $\epsilon_{j,j\oplus 2,l}$  is the completely antisymmetric Levi-Civita
tensor and $\oplus$ indicates addition modulo 4. Therefore, due to the one-to-one correspondence between the states $|u'_j\rangle$ and $|\psi_j\rangle$, if a click in the detector $j$ leads us to identify the input state as $|\psi_j\rangle$, the above equation tells us that our confidence in doing so will be $1/2$, which is the maximum one in the second stage of the SMC measurement [see Eq.~(\ref{eq:MC_1_2})]. 

When an inconclusive result is obtained from the two-outcome POVM in the second stage of the SMC measurement, the photon polarization in step VI is projected onto $|1\rangle_{\rm a}$. In this case, one can see from Eq.~(\ref{eq:xi_transf_exp}) that the states $|\xi_j\rangle$ are projected onto $|0\rangle$ for all $j=0,\ldots,3$. Therefore, a click in the photodetector ``?'' (box VIII in Fig.~\ref{fig:setup}) gives no information about the input states $|\psi_j\rangle$. The probability that this occurs after the two-stage SMC measurement will be, according with Eq.~(\ref{eq:prob_???_qutrit}), 
\begin{equation}
P_?^{\rm SMC} =P(?)P'(?)=c_0^2-c_1^2.
\end{equation}

To conclude this section, we would like to make two remarks: (i) The implementation of eight-port interferometers, as those sketched in boxes IV and VII of Fig.~\ref{fig:setup}, would be the most challenging steps in a possible realization of the optical scheme proposed here. However, this type of interferometer has been recently implemented for carrying out UD among three nonorthogonal states \cite{Mohseni04}, which ensures the feasibility of our scheme. (ii) The optical network proposed here can be, in principle, generalized for more states and higher dimensions. For $N$ symmetric states in a $D$-dimensional Hilbert space, the $D$ propagation modes are obtained by inserting $D$ HWPs and $D-1$ PBSs at the preparation step (box I in Fig.~\ref{fig:setup}). At each possible stage of the SMC measurement the two-outcome POVM is implemented with the polarization as ancilla. If it succeeds, a $2N$-port interferometer followed by photodetectors at each of its outputs will implement the $N$-outcome POVM that identifies the input states with maximum confidence in that particular stage.

\section{Conclusions}     \label{sec:Conc}
We have investigated a measurement strategy for discriminating among $N$ nonorthogonal symmetric qudit states with maximum confidence. Our study was restricted to a set of linearly dependent and equally likely pure states. For this problem, we found the optimal POVM that maximizes our confidence in identifying each state in the set and minimizes the probability of obtaining inconclusive results. The physical implementation of this POVM has been completely specified by considering the MC strategy as a two-step process \cite{Croke06,Croke08}. In the first step, a two-outcome POVM is performed with one outcome associated with the success and the other with the failure (or an inconclusive answer) of the process. To implement it, we introduced a two-dimensional ancilla and, in terms of the effect operators associated with each outcome, we prescribed the optimal unitary operation that provides the coupling between this ancilla and the original system. After measuring the ancilla, we showed that, in case of success, the input states are discriminated with maximum confidence in the second step of the MC strategy. This was achieved by applying an inverse discrete Fourier transform to the (transformed) input states and carrying out a projective measurement in the logical basis of an extended $N$-dimensional Hilbert space. On the other hand, in case of failure, it was shown that the input states can be mapped into a new set of equiprobable symmetric states, restricted to a subspace of the original qudit Hilbert space. As we discussed, if that was the case, the two-step MC measurement could be applied again onto this new set, and iterated in as many stages as allowed by the input states, until no further information could be extracted from an inconclusive result. We have shown that by implementing such optimized measurement, which we called ``sequential maximum-confidence measurement,'' our confidence in identifying the input states is the highest possible at each stage, although it decreases from one stage to the next. Also, the confidence per stage was shown to be higher than the one achieved by the optimal ME measurement if it had been applied in that stage.  For an $n$-stage SMC measurement we demonstrated that the probability of correctly identifying the input states increases the more stages we accomplish within the $n$ allowed. Finally, we have illustrated the SMC measurement in the simplest possible case where it can be applied, which is the discrimination among four qutrit states. For this particular case, we proposed an optical network, feasible with the current technology, which could carry out a two-stage SMC measurement and be generalized for more states and higher dimensions. 

It is important to remark that the SMC measurements for state discrimination may be, in principle, applied to an arbitrary set of linearly dependent qudit states. For the equally likely symmetric states studied here, the task to find the optimized POVM at each stage of the SMC measurement is facilitated, since the failure states are also symmetric and equiprobable. In the case of an arbitrary input set, this task will be certainly more complicated, since the form of the failure states and their associated probability distribution will not be necessarily the same at each stage. Fortunately, the MC strategy allows a closed form solution of the optimal POVM [Eq.~(\ref{eq:optimal_POVM})] for an arbitrary set of states.

The SMC measurement applied for symmetric qudit states, as proposed here, may have important applications. For instance, it is well known that after a unsuccessful attempt of unambiguously discriminating among equiprobable symmetric qudit states, the input set is mapped onto a linearly dependent set of states which are also symmetric and equally likely \cite{Chefles98-2,Jimenez07}. Due to the linear dependence, this set cannot be unambiguously discriminated by any further process. However, by carrying out a SMC measurement, our confidence in identifying the input states would significantly increase. We also anticipate other applications for the results presented in this article in the quantum communication protocols of entanglement swapping and quantum teleportation for high-dimensional quantum systems. When the quantum channel has nonmaximal Schmidt rank, it can be shown that these processes are mapped to the problem of discriminating linearly dependent symmetric states with the maximum confidence \cite{Jimenez11}. Moreover, for some quantum channels it might be possible to implement the SMC measurement introduced here, such that successful outcomes at each stage will lead to the highest possible fidelity (in that stage) for the protocol.

\begin{acknowledgments}
This work was supported by CONICYT PFB08-24, Milenio ICM P10-030-F, PBCT PDA-25, and FONDECYT 11085057. M.A.S.P. acknowledges financial support from CONICYT.
\end{acknowledgments}


\begin{thebibliography}{99}
\bibitem{CommentFourier} The external PSs are adjusted to compensate the relative phases arising from the reflections at the PBSs. By setting the internal PS to add a phase of $3\pi/2$, it can be shown that the eight-port interferometers in Fig.~\ref{fig:setup} implement the inverse Fourier transform ($\mathcal{\hat{F}}_4^{-1}$), where $\mathcal{\hat{F}}_4$ is given by Eq.~(\ref{eq:Fourier}) with $N=4$.
\end{thebibliography}
\end{document}